\documentclass[fleqn, usenatbib]{mnras}

\usepackage{newtxtext, newtxmath}

\usepackage[T1]{fontenc}

\usepackage{bm}

\DeclareRobustCommand{\VAN}[3]{#2}
\let\VANthebibliography\thebibliography
\def\thebibliography{\DeclareRobustCommand{\VAN}[3]{##3}\VANthebibliography}

\usepackage{graphicx}       
\usepackage{subcaption}     
\usepackage{amsmath}        
\usepackage[table]{xcolor}  
\usepackage{orcidlink}      
\usepackage{enumerate}      
\usepackage{enumitem}       

\usepackage{xspace}
\newcommand\LISA{{\it LISA}\xspace}

\title[Eccentric Galactic binaries in \LISA]{
    Discovering neutron stars with LISA via measurements of orbital eccentricity in Galactic binaries
}

\author[C.\ J.\ Moore et al.]{
    \!Christopher J. Moore$^{\orcidlink{0000-0002-2527-0213}}$$^{1,2,3}$\thanks{E-mail: cmoore@star.sr.bham.ac.uk},
    Eliot Finch$^{\orcidlink{0000-0002-1993-4263}}$$^{5,4}$,
    Antoine Klein$^{\orcidlink{0000-0001-5438-9152}}$$^{4}$,
    Valeriya Korol$^{\orcidlink{0000-0002-6725-5935}}$$^{6,4}$,
    Nhat Pham$^{\orcidlink{0009-0001-6389-4334}}$$^{4,7}$,
    \newauthor Daniel Robins$^{\orcidlink{0009-0007-2318-258X}}$$^{4,8}$
    \\
    $^{1}$Institute of Astronomy, University of Cambridge, Madingley Road, Cambridge, CB3 0HA, UK \\
    $^{2}$Kavli Institute for Cosmology, University of Cambridge, Madingley Road, Cambridge, CB3 0HA, UK \\
    $^{3}$Department of Applied Mathematics and Theoretical Physics, Centre for Mathematical Sciences, University of Cambridge, Wilberforce Road, CB3 0WA, UK \\
    $^{4}$Institute for Gravitational Wave Astronomy \& School of Physics and Astronomy, University of Birmingham, Birmingham, B15 2TT, UK \\
    $^{5}$TAPIR, California Institute of Technology, Pasadena, CA 91125, USA\\
    $^{6}$Max-Planck-Institut f{\"u}r Astrophysik, Karl-Schwarzschild-Stra{\ss}e 1, 85741 Garching, Germany\\
    $^{7}$School of Physics, HH Wills Physics Laboratory, Tyndall Avenue, Bristol, BS8 1TL, UK \\
    $^{8}$Department of Physics, University of Warwick, Gibbet Hill Road, Coventry, CV4 7AL, UK
}

\date{Accepted XXX. Received YYY; in original form ZZZ}

\pubyear{2023}

\begin{document}
\label{firstpage}
\pagerange{\pageref{firstpage}--\pageref{lastpage}}
\maketitle

\begin{abstract}
    \LISA will detect $\sim \! 10^4$ Galactic binaries, the majority being double white dwarfs. 
    However, approximately $\sim \! 1 \textrm{--} 5 \%$ of these systems will contain neutron stars which, if they can be correctly identified, will provide new opportunities for studying binary evolution pathways involving mass reversal and supernovae as well as being promising targets for multi-messenger observations.
    Eccentricity, expected from neutron star natal kicks, will be a key identifying signature for binaries containing a neutron star.
    Eccentric binaries radiate at widely-spaced frequency harmonics that must first be identified as originating from a single source and then analysed coherently. 
    A multi-harmonic heterodyning approach for this type of data analysis is used to perform Bayesian parameter estimation on a range of simulated eccentric \LISA signals. 
    This is used to: 
    (i) investigate \LISA's ability to measure orbital eccentricity and to quantify the minimum detectable eccentricity;
    (ii) demonstrate how eccentricity and periastron precession help to break the mass degeneracy allowing the individual component masses to be inferred, potentially confirming the presence of a neutron star;
    (iii) investigate the possibility of source misidentification when the individual harmonics of an eccentric binary masquerade as separate circular binaries;
    and (iv) investigate the possibility of source reclassification, where parameter estimation results of multiple circular analyses are combined in postprocessing to quickly infer the parameters of an eccentric source.
    The broader implications of this for the ongoing design of the \LISA global fit are also discussed.
\end{abstract}

\begin{keywords}
    gravitational waves -- 
    binaries: close -- 
    stars: white dwarfs -- 
    stars: neutron
\end{keywords}


\section{Introduction}\label{sec:introduction}

The Laser Interferometer Space Antenna \citep[\LISA;][]{2017arXiv170200786A} is a planned space-based gravitational-wave (GW) observatory being developed for launch in the mid 2030s. 
\LISA will observe the source-rich mHz GW spectrum, which is populated by $\sim 10^8$ short-period ($\sim \!1\,\mathrm{hour}$) stellar-mass compact-object binaries in our galaxy \citep{1987A&A...176L...1L, 1990ApJ...360...75H, 2001A&A...375..890N} and its satellites \citep{2020ApJ...894L..15R, 2021MNRAS.502L..55K}. 
These weak-field GW sources generate quasi-monochromatic signals and will be the most numerous \LISA sources.
The majority will be double white-dwarf (WD+WD) binaries; in addition to a few tens of known \emph{verification} systems \citep[see, e.g.,][]{2023arXiv230212719K, 2023MNRAS.522.5358F} \LISA is expected to discover a few tens of thousands of individually detectable WD+WD binaries \citep{2004MNRAS.349..181N, 2012ApJ...758..131N, 2017ApJ...846...95K, 2020ApJ...898...71B, 2020ApJ...893....2L, 2020A&A...638A.153K, 2022MNRAS.511.5936K} and to be able to probe the stochastic GW signal generated by the entire Galactic population of several million binaries \citep{1990ApJ...360...75H, 1997CQGra..14.1439B, 2003MNRAS.346.1197F, 2010ApJ...717.1006R, 2023MNRAS.519.2552G}. 

\begin{figure*}
    \centering
    \includegraphics[width=0.98\textwidth]{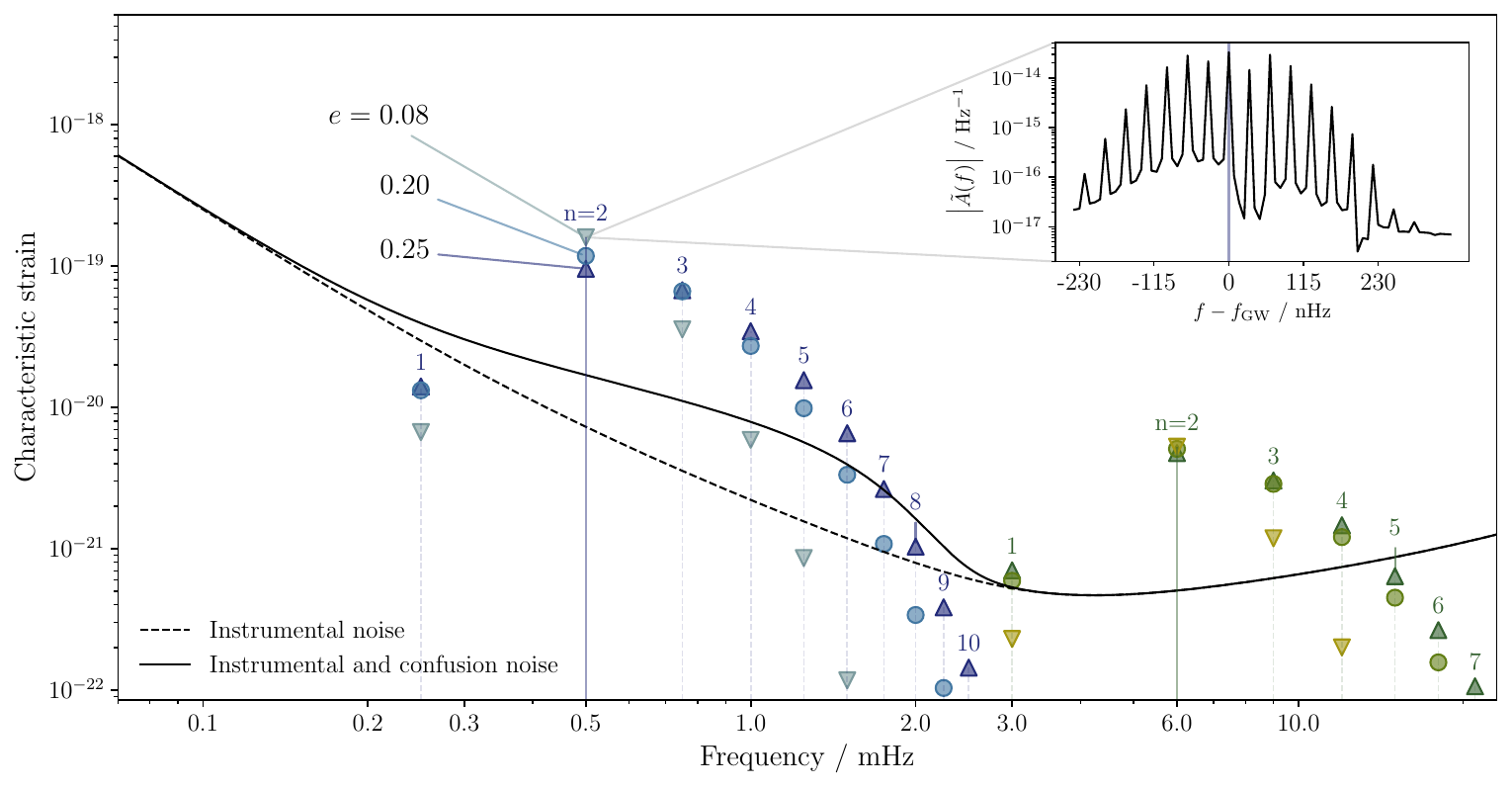}
    \caption{ \label{fig:freq:ecc}
        The harmonic frequency structure for an illustrative set of six eccentric quasi-monochromatic binaries.
        Each binary is either low frequency ($f_{\rm GW}=0.5\,\mathrm{mHz}$, blue shades) or high frequency ($f_{\rm GW}=6\,\mathrm{mHz}$, green shades) and either low ($e=0.08$, down-pointing triangles), medium ($e=0.2$, circles), or high ($0.25$, up-pointing triangles) eccentricity.
        Harmonics arise at half-integer multiples of the frequency of the dominant $n=2$ harmonic.
        Also shown is the \LISA noise curve, with and without the expected contribution from the Galactic confusion noise for a 4 year mission. 
        For each harmonic, the height of the marker relative to the noise curve indicates the SNR in that harmonic.
        The harmonics are extremely narrow band; the inset plot shows the structure of one harmonic in one \LISA TDI channel. 
        The spikes clearly visible in the inset are spaced at frequency intervals of $1\,\mathrm{year}^{-1}$ and are due to the orbital modulation of the instrument response. 
        All sources were simulated for a 4-year \LISA mission, with $\dot{f}_{\rm GW} = 5.99\times 10^{-19} \,\mathrm{Hz}\,\mathrm{s}^{-1}$, $f_{\rm PP}=3.91\,\mathrm{nHz}$, $\iota=\pi/4$, and in the direction of the Galactic centre (ecliptic coordinates of $[l,b] = [4.66, -0.098]$ radians). 
        The total SNR (summed across all harmonics) of each source was chosen to be $\rho = 10$.
    }
\end{figure*}

\LISA will also detect Galactic quasi-monochromatic binaries including a neutron star (NS) \citep{2023IAUS..363..203K} and/or black hole \citep{2022ApJ...937..118W} component, albeit in far smaller numbers. 
Of particular interest here are the several hundred neutron star - white dwarf (NS+WD) binaries \citep{2020ApJ...898...71B, 2024MNRAS.530..844K} that are now expected to be detectable by \LISA \citep[this is significantly more than early estimates,][]{2001A&A...365..491N}. 
Formation of these sources is expected from isolated binaries evolving in the field with a significant fraction of binaries undergoing mass reversal \citep{2018A&A...619A..53T, 1993MNRAS.260..675T, 2000A&A...355..236T, 2002MNRAS.335..369D, 2017MNRAS.467.3556B, 2019MNRAS.486.1805Z} during a period of stable mass transfer which results in the WD forming before the NS. 
Then, when the NS forms, the accompanying supernovae kick puts the binary onto an eccentric orbit. 
This picture is supported by observations of several binary-pulsar systems with WDs on eccentric orbits \citep{1989Natur.340..367L, 1999ApJ...516L..25V, 2000ApJ...543..321K, 
2001ApJ...547..345B, 2018MNRAS.476.4315N}.

Observing NS+WD binaries with \LISA will help in understanding the mass transfer and supernovae kicks in their evolution as well as providing new targets for multi-messenger astronomy \citep{2023IAUS..363..203K}. 
An important question is if it will be possible to tell if a binary contains a NS from its GW signal alone? 
This is challenging because quasi-monochromatic GW signals contain very little information. 
For exactly monochromatic binaries, the binary masses cannot be inferred at all, and even if the binary is chirping (and assuming this is due to GW emission), only the chirp mass combination can be inferred \citep{2021MNRAS.502.5576K}.
Orbital eccentricity might be an important \emph{smoking gun}; WD+WD binaries are expected to have been efficiently circularised by several episodes of mass transfer, so measuring a non-zero eccentricity strongly suggests a NS+WD source, or another rare type of Galactic binary (e.g., involving a black hole) or a WD+WD binary from an unusual evolutionary pathway.
Furthermore, for binaries on eccentric orbits, the rate of advance of the periapsis in general relativity (GR) depends on the total mass of the binary; therefore, if the periastron precession (also known as apsidal precession) frequency can be measured from the GW signal then this allows the total mass of the binary to be inferred \citep[][see also Eq.~\ref{eq:periastron_prec_freq}]{2001PhRvL..87y1101S}. 
For binaries where the chirp and periapsis precession can both be measured, it will be possible to infer the total mass and the chirp mass separately, breaking the degeneracy and allowing the individual component masses to be determined. 

The effects of orbital eccentricity imprint themselves on the GW signals of quasi-monochromatic binaries in several ways \citep{1963PhRv..131..435P, 1964PhRv..136.1224P, AIHPA_1985__43_1_107_0, 1994MNRAS.266...16M, 1995MNRAS.274..115M, 2001PhRvL..87y1101S, 2001PhRvD..63d4023D, 2005PhRvD..72b9902D, 2015PhRvD..91h4040M}.
The most striking feature is the presence of narrow-band harmonics at multiples of the orbital period, $P$, with frequencies $f_n\sim n/P$ (see, e.g., Fig.~\ref{fig:freq:ecc}).
The relativistic effect of periastron precession (cf.\ the orbit of Mercury) causes the orientation of the orbital ellipse to advance over time and precess with a frequency $f_\mathrm{PP} \ll f_{n=2}$. 
This periastron precession modulates the GW signal, further shifting the frequency of the harmonics by multiples of $f_\mathrm{PP}$ (an expression for the frequency of the $n^{\rm th}$ harmonic is given in Eq.~\ref{eq:circ_ecc_f}).
Finally, orbital eccentricity also enhances the overall flux of GWs, leading to a quicker inspiral and the frequency of each harmonic increasing faster with time.

Narrow-band signals can be efficiently analysed using a heterodyning approach focusing on the small number of relevant frequency bins. 
This paper uses a multi-harmonic heterodyning algorithm for the efficient analysis of multiple narrow bands widely separated in frequency.
A new problem of source confusion also arises when analysing eccentric signals. 
The multiple, narrow frequency harmonics generated by an eccentric source can be mistaken for separate circular sources. 
It is possible to accurately match an eccentric GW source with a series of circular templates (one for each harmonic), thereby mistaking one source for several. 
This has implications for \LISA data analysis where the numerous sources in the millihertz band generate GW signals that are overlapping in both time a frequency space and must all be analysed together in a process called a global fit.
Different parts of the global fit algorithm designed to target different source types (and which may need to be run asynchronously) must be tolerant to mistaken source identifications made by other parts of the algorithm. 
Different parts of the algorithm will need to share information to find and correct such mistakes, e.g.\ by reclassifying multiple circular source candidates as a single eccentric source. 

The structure of this paper is as follows:
The basic physics of eccentric Galactic binaries and their GW signals is described in Sec.~\ref{sec:physics}.
The data analysis methods used are described in Sec.~\ref{sec:heterodyning}.
The three main results of this paper are presented in Sec.~\ref{sec:res}:
Sec.~\ref{sec:results} quantifies \LISA's minimum measurable eccentricity for quasi-monochromatic Galactic binaries as a function of source parameters; 
Sec.~\ref{sec:comp_masses} verifies that for certain systems, eccentricity allows the component masses to be inferred, potentially confirming the presence of a NS; 
Sec.~\ref{sec:results_confusion} considers the problem of source misidentification;
and Sec.~\ref{subsec:reclassify} considers the problem of source reclassification.
Finally, Sec.~\ref{sec:global fit} concludes by discussing some of the broader implications of this work for ongoing work on designing the \LISA global fit.

\section{GW physics of eccentric binaries}\label{sec:physics}

Quasi-monochromatic Galactic binaries are weak field sources of GWs.
Therefore, the GW signals can be computed (in the transverse-traceless gauge) using the quadrupole formula \citep{1973grav.book.....M}.
This has been investigated for a binary on a Keplerian elliptical orbit by several authors \citep[see, e.g.,][]{1995MNRAS.274..115M, 2001PhRvL..87y1101S}.
We employ the following expressions for the two polarizations:
\begin{align}\label{eq:GW}
    h_+ &= -(1 + \cos^2{\iota}) \sum_{n=1}^N \mathcal{A}_n \cos[2\pi f_n t + \phi + (n-2)\Phi] \nonumber \\
    h_\times &= 2\cos{\iota} \sum_{n=1}^N \mathcal{A}_n \sin[2\pi f_n t + \phi + (n-2)\Phi].
\end{align}
This signal contains discrete harmonics at frequencies $f_n$, with corrections coming from the periastron precession.
Here, $\mathcal{A}_n(e)$ is the amplitude of each harmonic, 
$\iota$ is the inclination,
$\phi$ is the phase,
$\Phi$ is the argument of periastron, 
and $e$ is the eccentricity.

Typically, the number of harmonics $N$ included should be chosen depending on the eccentricity and the frequency in order to capture a given fraction of the squared signal-to-noise ratio (SNR; see Eq.~\ref{eq:snr}).
However, to facilitate comparisons between sources with different eccentricities, most of the calculations in this paper were performed with a fixed value $N=12$. 
For the small to moderate eccentricities considered in this paper, this is a conservative choice as there is negligible power at larger $n$.
The effect of doing the analysis with different numbers of harmonics is investigated in Appendix \ref{app:varyN}.

\begin{figure}
    \centering
    \includegraphics[width=0.98\columnwidth]{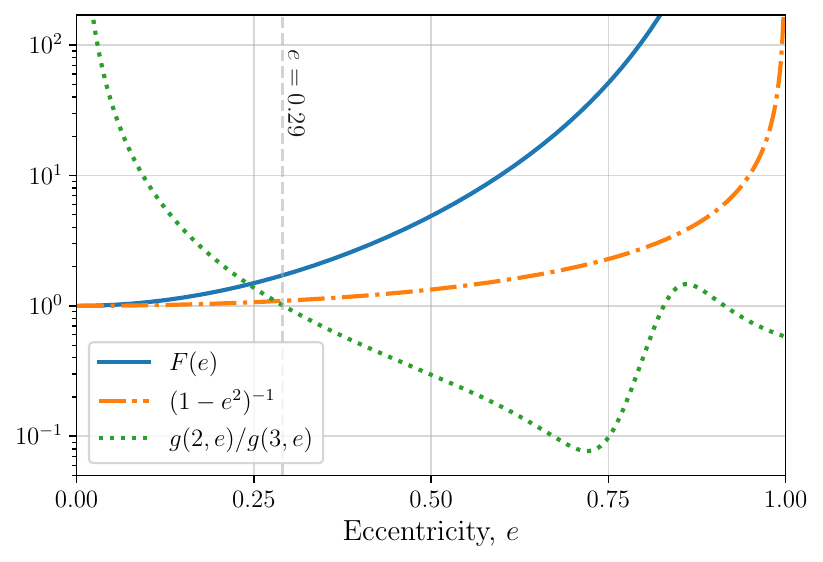}
    \caption{
        Plots of various eccentricity dependent quantities that enter the waveform. The factor $F(e)$ describes the enhancement of the chirp rate compared to a circular binary with the same frequency.
        The quantity $(1-e^2)^{-1}$ enters the periastron precession frequency (Eq.~\ref{eq:periastron_prec_freq}) and the GW amplitude (Eq.~\ref{eq:amp_analytical}).
        Finally, $g(2,e)/g(3,e)$ is the ratio of the power in the $n=2$ and 3 harmonics \citep[the notation $g(n,e)$ is from][]{1963PhRv..131..435P} which dominate for small eccentricities; when $e<0.29$, the $n=2$ contains the most power. At very high eccentricities ($e\gtrsim 0.7$) this ratio increases again, however by this point the power is spread over many harmonics with $n\gg 3$.
    }
    \label{fig:ecc_functions}
\end{figure}

Instead of the orbital period $P$, it will be more convenient for us to work with the initial frequency of what is (usually, at least when $e\lesssim 0.29$; see Fig.~\ref{fig:ecc_functions}) the dominant harmonic, $f_{\rm GW} = f_{n=2}(t=0) = 2/P$.
(The time $t=0$ is defined as the start of \LISA science operations.)
A binary radiating GWs does not stay on a fixed Keplerian orbit and to account for this the frequency is allowed to drift, or chirp, linearly in time according to 
\begin{align}\label{eq:chirprate}
    f_n = \frac{n}{2}(f_{\rm GW}+ \frac{1}{2}\dot{f}_{\rm GW}t) .
\end{align}
The argument of periastron is also allowed to advance according to 
\begin{align}\label{eq:prec}
    \Phi=\Phi_0+2\pi f_{\rm PP}t,
\end{align}
where $f_{\rm PP}$ is the periastron precession frequency.

Note that we use different conventions than those in the references, which explains the different form of our polarizations.
For example, when comparing with \citet{1995MNRAS.274..115M}, we only keep the dominant $(S_n + C_n)/2$ terms \citep[where $S_n$ and $C_n$ are functions of $e$ involving the Bessel functions of the first kind $J_n$; see Eqs.~A15 and A16 of][]{1995MNRAS.274..115M}.
We also choose to give the frequency of each harmonic in terms of the periastron-to-periastron frequency, $f_\mathrm{peri}$ (such that $f_\mathrm{GW} = 2f_\mathrm{peri}$), whereas the references give frequencies relative to the orbital frequency $f_\mathrm{orb}$ (that is, the frequency of sweeping out $2\pi$ radians).
These are related via $f_\mathrm{orb} = f_\mathrm{peri} + f_\mathrm{PP}$.

Our GW polarization amplitudes, $\mathcal{A}_n$, are implemented via an expansion correct up to tenth order in eccentricity.
While this approach is obviously not suitable for eccentricities close to unity, it has been found to work extremely well for the small to moderate eccentricities $e\lesssim 0.6$ expected in WD+NS binaries.
To make the connection to \citet{1995MNRAS.274..115M}, the harmonic amplitudes are the expansions of
\begin{equation}\label{eq:amp_analytical}
    \mathcal{A}_n = \mathcal{A} \frac{S_n(e) + C_n(e)}{2(1-e^2)} 
\end{equation}
where $\mathcal{A}$ is an overall amplitude common to each harmonic (see Eq.~\ref{eq:GWamp}).
For low values of eccentricity, we can also write
\begin{equation}\label{eq:circ_ecc_A}
    \mathcal{A}_n \approx \mathcal{A} \frac{2}{n} g(n,e)^{1/2}
\end{equation}
where $g(n,e)$ is a function proportional to the power in the $n^\mathrm{th}$ harmonic \citep[notation from][]{1963PhRv..131..435P}.
When expanded about $e=0$, Eqs.~\ref{eq:amp_analytical} and \ref{eq:circ_ecc_A} are equal up to and including the $(n+1)^\mathrm{th}$ power in $e$ (for $n \geq 2$).

The \LISA response to GWs must also be modelled. 
This depends on the sky position (ecliptic coordinates $l$ and $b$) and orientation (via the polarisation angle $\psi$) of the source.
The orbital motion of \LISA modulates the response at a frequency of $1\,\mathrm{year}^{-1}$ and introduces Doppler broadening into each frequency mode.
The \LISA phase measurements are processed into three time series containing the GW information through a process known as time-delay interferometry \citep[TDI;][]{2021LRR....24....1T} which are designed to suppresses the large-amplitude laser noise below the level of the other noise sources.
The model for the \LISA response is described in \cite{2021arXiv210610291K}.
Schematically,
\begin{align} \label{eq:LISAresponse}
    A, E, T = \mathrm{LISA}\_\mathrm{Response}(h_+,h_\times;\,l,b,\psi),
\end{align}
where $A$, $E$ and $T$ are the noise-orthogonal TDI output channels. 
The log-likelihood is expressed in terms of these time series (see Eq.~\ref{eq:like}).

Eqs.~\ref{eq:GW} to \ref{eq:LISAresponse} constitute our GW waveform model for eccentric quasi-monochromatic binaries. 
There are 11 independent signal parameters: $\{\mathcal{A}, \iota, f_{\rm GW}, \dot{f}_{\rm GW}, f_{\rm PP}, e, \phi, \Phi_0, l, b, \psi\}$.
When working with the cosine of the inclination and the sine of the ecliptic latitude, inference was performed directly on these parameters with uniform priors. 
We chose to allow the parameter inference to sample values of the parameters that would be nonphysical if the binary was simply two point masses moving under the influence of gravity alone (for example, we are free to use priors that allow negative $\dot{f}_{\rm GW}$ or $f_{\rm PP}$). 
This is done to allow the analysis to detect binaries that may be interacting, e.g.\ via mass transfer or tidal forces.
The conversion to the physical binary parameters described below is performed, if required, in post processing.
It should be remembered that these conversions will give biased results (or may fail entirely) if the binary is interacting.

Consider a binary with masses $m_1$ and $m_2$ and semi-major axis $a$.
The orbital period is related to $a$ and the total mass $M=m_1+m_2$ by Kepler's third law; in terms of $f_{\rm GW}$, this is 
\begin{align} \label{eq:KeplerIII}
    f_{\rm GW} &= \frac{1}{\pi} \sqrt{\frac{GM}{a^3}}\\ &= \left(\frac{M}{M_\odot}\right)^{1/2}\left(\frac{a}{R_\odot}\right)^{-3/2} \times 0.20\,\mathrm{mHz} .
\end{align}
As the binary inspirals the frequency increases, or chirps, with time
\begin{align} \label{eq:chirp}
    \dot{f}_{\rm GW} &= \frac{96}{5\pi c^5}(G\mathcal{M}_c)^{5/3}(\pi f_{\rm GW})^{11/3}F(e)\\ &= \left(\frac{\mathcal{M}_c}{M_\odot}\right)^{5/3}\left(\frac{f_{\rm GW}}{1\,\mathrm{mHz}}\right)^{11/3} F(e) \times 0.18\,\mathrm{nHz}\,\mathrm{year}^{-1}.
\end{align}
Here, an overdot denotes differentiation with respect to time, $\mathcal{M}_c=(m_1m_2)^{3/5}/(m_1+m_2)^{1/5}$ is the binary chirp mass, and 
\begin{align} \label{eq:enhancement_factor}
    F(e) = \frac{1+\frac{73}{24}e^2 + \frac{37}{96}e^4}{(1-e^2)^{7/2}} 
\end{align}
is the eccentricity enhancement \citep[see Fig.~\ref{fig:ecc_functions};][]{1963PhRv..131..435P}.
At leading post-Newtonian order, the argument of periastron advances with time.
The periastron precession frequency is
\begin{align} \label{eq:periastron_prec_freq}
    f_{\rm PP} &= \frac{3}{2 \pi c^2}\frac{(GM)^{2/3}}{(1-e^2)}(\pi f_{\rm GW})^{5/3}\\ &= \frac{1}{1-e^2}\left(\frac{M}{M_\odot}\right)^{2/3}\left(\frac{f_{\rm GW}}{1\,\mathrm{mHz}}\right)^{5/3} \times 0.29\,\mathrm{year}^{-1}.
\end{align}
The frequency derivative depends on $\mathcal{M}_c$ while the periastron advance depends on $M$. 
Therefore, if both $\dot{f}_{\rm GW}$ and $f_{\rm PP}$ can be measured, it is possible to determine $m_1$ and $m_2$ independently.
Finally, if the source is at a distance $D$, the GW strain amplitude is given by
\begin{align} \label{eq:GWamp}
    \mathcal{A} &=  \frac{2(G\mathcal{M}_c)^{5/3}(\pi f_\mathrm{GW})^{2/3}}{D c^4} \\
    &= \left(\frac{\mathcal{M}_c}{M_\odot}\right)^{5/3} \left(\frac{f_\mathrm{GW}}{1\,\mathrm{mHz}}\right)^{2/3} \left(\frac{10\,\mathrm{kpc}}{D}\right) \times 5.94\times 10^{-23} .
\end{align}

\begin{figure*}
    \centering
    \includegraphics[width=0.98\textwidth]{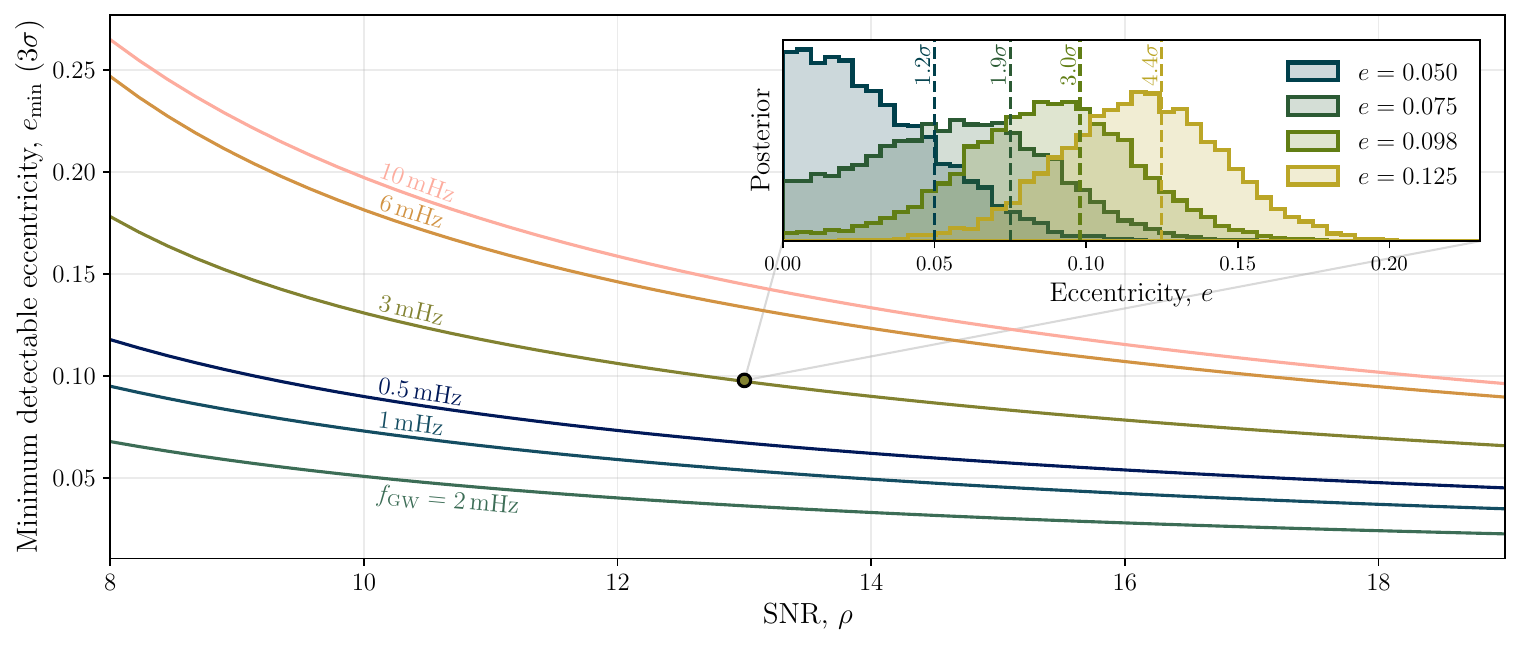}
    \caption{
    The minimum detectable eccentricity as a function of SNR, for a selection of the GW frequencies tested.
    To a good approximation, the minimum eccentricity of a quasi-monochromatic binary that \LISA can measure depends only on the frequency of the source and its SNR.
    These curves were constructed using a 4-year \LISA mission, with source properties corresponding to a $1.4\,M_\odot\text{--}1.4\,M_\odot$ binary in the direction of the Galactic centre, at an inclination $\iota = \pi/4$; however, as verified in Appendix \ref{app:vary_Tobs}, the results are not sensitive to these choices.
    The inset shows a sequence of eccentricity posteriors for a $3\,\mathrm{mHz}$ source at an SNR of 13, with four different injected eccentricities.
    The significance at which each eccentricity posterior is peaked away from zero (the median divided by the standard deviation of the distribution) is indicated along the vertical lines (which are placed at the injected value of the eccentricity).
    This shows visually what is meant by the threshold criterion of requiring the median over standard deviation to exceed 3 before claiming a eccentricity detectable. 
    These results are confirmed using instead a threshold on the evidence ratio (or Bayes' factor) in Appendix \ref{app:BF}.
    }
    \label{fig:min_ecc}
\end{figure*}

\begin{figure}
    \centering
    \includegraphics[width=0.97\columnwidth]{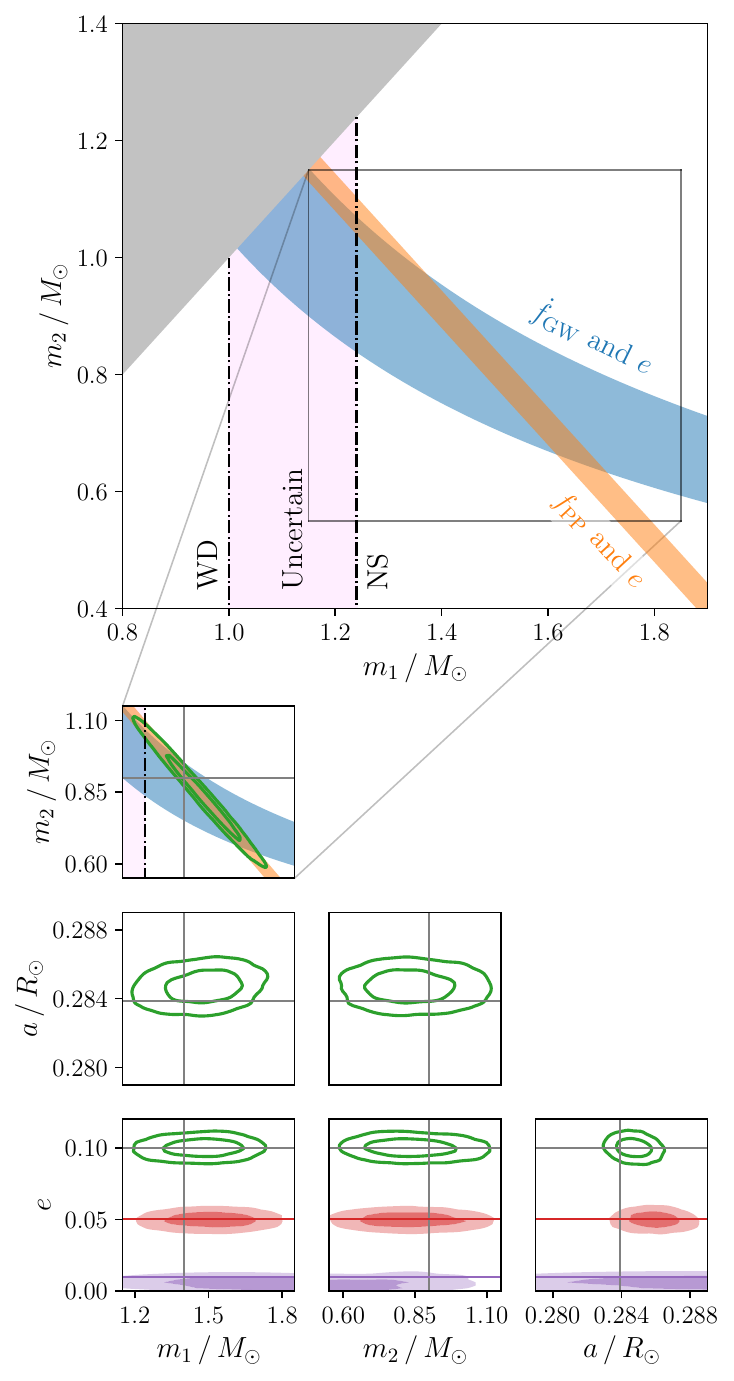}
    \caption{
        \LISA parameter estimation results for a highly eccentric WD+NS binary (green contours).
        The simulated binary parameters were: $m_1=1.4\,M_\odot$ and $m_2=0.9\,M_\odot$, $P=16.7\,\mathrm{minutes}$ (meaning $f_{\rm GW}=2\,\mathrm{mHz}$), $e=0.1$, $\iota=\pi/3$, $D=10\,\mathrm{kpc}$, and in the direction of the Galactic centre. 
        The simulated binary is detached, meaning $\dot{f}_{\rm GW}$ and $f_{\rm PP}$ take the GR values given by Eqs.~\ref{eq:chirp} and \ref{eq:periastron_prec_freq} respectively.
        These parameter choices give a total source SNR of $\rho=34.8$ after 4 years of observation. 
        Joint 2-dimensional posterior distributions are shown on the parameters $m_1$, $m_2$, $a$ and $e$. 
        (The convention $m_1 \geq m_2$ excludes the grey region of parameter space.)
        The top panel shows how the measurements of the chirp mass (coming from the chirp rate $\dot{f}_{\rm GW}$) and the total mass (coming from the periastron precession $f_{\rm PP}$) combine to break the mass degeneracy and allow both $m_1$ and $m_2$ to be determined separately.
        Vertical dot-dashed lines indicate the approximate upper and lower ends of the WD and NS mass distributions in a particular population synthesis model (described in the text) and show that from the mass measurements (together with that the binary is eccentric; $e=0.1\pm 0.005$) we could confidently conclude that this binary contains a NS from the \LISA observation of its GW signal alone.
        The bottom row shows selected posteriors for similar binaries with smaller eccentricities of $e=0.05$ (red) and $e=0.01$ (purple), with SNRs 29.3 and 27.2 respectively. 
        (When $e=0.01$, the measured eccentricity is consistent with zero and it is no longer possible to determine the component masses.)
    }
    \label{fig:golden_binary}
\end{figure}

\section{\LISA Data analysis with multiple harmonics}\label{sec:heterodyning}

The most important ingredient for Bayesian GW data analysis is the likelihood function. 
Under the common simplifying assumptions that the noise in \LISA has a known power spectral density (PSD) and is additive, Gaussian and independent in each of the three noise-orthogonal TDI channels, the likelihood is given by \begin{align} \label{eq:like}
    P(d|\theta) \propto \prod_\alpha \exp\left(-\frac{1}{2}\big<d-h(\theta)\big|d-h(\theta)\big>_\alpha\right) .
\end{align}
Here, $d$ is the data, $h(\theta)$ is the waveform model as a function of the 11 source parameters, $\theta$, and angled brackets denote the signal inner product in each of the three TDI channels, $\alpha\in\{A,E,T\}$.
The factors not shown in the proportional relation in Eq.~\ref{eq:like} are unimportant constants in the sense that they do not depend on $\theta$.

The inner products are defined as integrals (or sums for discretely sampled data) over frequency.
However, the long \LISA mission and accompanying fine frequency resolution ($\Delta f = 1/T_{\rm obs}$) coupled with quasi-monochromatic nature of these sources mean that it would be extremely inefficient to sum over all frequencies. 
Instead, the signal inner product in Eq.~\ref{eq:like} is split into a sum of integrals targeting narrow frequency ranges around each harmonic;
\begin{align}  
    \big<\tilde{x}(f)\big|\tilde{y}(f)\big>_\alpha = \sum_{n=1}^N \big<\tilde{x}(f)\big|\tilde{y}(f)\big>^n_\alpha ,
\end{align}
where 
\begin{align} \label{eq:narrow_sum}
    \big<\tilde{x}(f)\big|\tilde{y}(f)\big>^n_\alpha = 4\mathrm{Re}\bigg\{\int_{f_{n,\mathrm{low}}}^{f_{n,\mathrm{high}}} \mathrm{d}f \; \frac{\tilde{x}_\alpha(f) \tilde{y}^*_\alpha(f)}{S_\alpha(f)}\bigg\}.
\end{align}
A subscript $\alpha$ denotes the time series in the relevant TDI channel and a star denotes complex conjugation.
The integration limits $(f_{n,\mathrm{low}}, f_{n,\mathrm{high}})$ are chosen to target each frequency harmonic based on the source priors, and the bandwidths $|f_{n,\mathrm{high}}-f_{n,\mathrm{low}}|$ are chosen to be wide enough to capture all the signal power in a given harmonic after accounting for the source chirp and Doppler broadening caused by the motion of the \LISA constellation, 
while avoiding spectral leakage between different harmonics; this can occur when negative frequency components of one harmonic get heterodyned onto another harmonic.
For example, $f_{n,\mathrm{high}}$ is determined by taking the upper end of the prior ranges for $f_{\rm GW}$ and $f_{\rm PP}$ combined using Eq.~\ref{eq:circ_ecc_f}, plus a term depending on the upper end of the prior range for $\dot{f}_{\rm GW}$ calculated using Eq.~\ref{eq:circ_ecc_fdot}, plus a term corresponding to the maximum possible Doppler shift if the source happens to lie in the ecliptic plane.

Each of the narrow-band sums in Eq.~\ref{eq:narrow_sum} can then be efficiently evaluated by using heterodyning to
shift the frequency range of interest close to zero, low-pass filtering the data and down sampling to target only the small number of frequency bins in the range $f_{n,\mathrm{low}}$ to $f_{n,\mathrm{high}}$ (with several extra padding bins included for safety).

An important quantity in GW data analysis is the (optimum) SNR.
This largely determines if a source is detectable and any parameter estimation uncertainties.
The SNR is defined as
\begin{align} \label{eq:snr}
    \rho^2= \sum_\alpha \big<h|h\big>_\alpha .    
\end{align}
The total source SNR is the quadrature sum of the SNRs in the individual harmonics; i.e.\ $\rho^2=\sum_{n=1}^{N}\rho_n^2$, where
\begin{align} \label{eq:snr_n}
    \rho^2_n = \sum_\alpha \big<h\big|h\big>^n_\alpha .
\end{align}

The inner products in Eq.~\ref{eq:like} were evaluated using a simple analytic model for the noise PSD based on the latest \LISA science requirements (SciRD). 
This model includes estimates of the contributions from both the instrumental noise \citep[following][]{2021_Babak_PSD} and the Galactic confusion noise \citep[using Eq.~4 of][scaled to the desired mission duration]{Babak_EMRI_2017}. \cite{2024MNRAS.530..844K} have verified that the population of eccentric sources (not present in Babak et al.) does not contribute to the confusion foreground.

The analysis was performed using the \textsc{Balrog} software package that is being developed for \LISA data analysis and parameter estimation for all source types. This includes supermassive binary black hole mergers \cite{2023PhRvD.107l3026P}, double WDs \citep{2019PhRvD.100h4041B,2020ApJ...894L..15R,2023MNRAS.522.5358F}, and stellar-mass binary black holes \citep{2021PhRvD.104d4065B,2022arXiv220403423K, 2023arXiv230518048B}. 

Parameter estimation was performed using the \textsc{nessai} implementation \citep{nessai, 2021PhRvD.103j3006W} of the nested sampling algorithm \citep{2004AIPC..735..395S, Skilling:2006gxv}.
Sampling was performed directly on the 11 parameters described in Sec.~\ref{sec:physics} using flat priors.

\section{Results} \label{sec:res}

\subsection{The Minimum Detectable Eccentricity}\label{sec:results}

This section aims to establish the minimum eccentricity, $e_\mathrm{min}$, that a quasi-monochromatic binary must have in order for \LISA to be able to measure it. 
In other words, the smallest eccentricity that allows us to confidently conclude a binary is not circular. 
It is expected that $e_\mathrm{min}$ will depend strongly on $f_\mathrm{GW}$ and the SNR, $\rho$, as these directly control the strength of the various harmonics relative to the \LISA noise curve. 
The influence of the other parameters is expected to be smaller and/or tend to average out after several orbits of the \LISA constellation; this is verified in Appendix \ref{app:vary_Tobs}.

The analysis framework described in Sec.~\ref{sec:heterodyning} was used to perform a series of zero-noise injections with different values of $f_\mathrm{GW}$, $\rho$ (which was controlled by varying $\mathcal{A}$), and $e$.
Sources were simulated at 13 values of SNR, $\rho\in\{7,8,\ldots,19\}$, and 16 values of frequency, $f_{\rm GW}/\mathrm{mHz}\in\big\{0.5, 0.75, \ldots, 2.75, 3,4,5,6,8,10\}$.
For each pair of $f_\mathrm{GW}$ and $\rho$, a series of analyses were performed at different $e$ to find $e_{\rm min}$. 
All other source parameters were held constant at the values described in the caption of Fig.~\ref{fig:min_ecc}. 

The significance of the eccentricity measurement was quantified via the ratio of the median to the standard deviation, denoted $\sigma$, of the 1-dimensional marginal posterior on $e$ (see inset plot in Fig.~\ref{fig:min_ecc}).
For simplicity, the eccentricity is considered detectable if this ratio exceeds a fixed threshold value of 3. 
These results were also checked using the Bayes' factor between an analysis using circular and eccentric waveform models with similar results (see Appendix \ref{app:BF}).

The minimum detectable eccentricity is plotted in Fig.~\ref{fig:min_ecc}.
For convenience, a fitting formula for these results is also provided;
\begin{align} \label{eq:analytic_emin}
    e_\mathrm{min}(f_\mathrm{GW}, \rho) \approx &\left(\frac{1}{\rho^{1.54}} + \frac{1}{\rho} \right) \nonumber \\
    \times \{1.08 &+ 0.87\tan^{-1}\left[1.08 \left(f_\mathrm{GW}/\mathrm{mHz} - 2.13 \right)\right] \nonumber \\
    &- 0.55\tan^{-1}\left[2.08 \left(f_\mathrm{GW}/\mathrm{mHz} - 1.22 \right)\right]\}. 
\end{align}
This fit has been tested for $0.5 \leq f_{\rm GW}/\mathrm{mHz} \leq 10$ and $\rho \geq 8$.
For large $\rho$, the fit scales as $e_{\rm min}\propto\rho^{-1}$, so Eq.~\ref{eq:analytic_emin} can be extrapolated to high SNRs.
However, this formula should not be extrapolated to lower SNRs, where the $\rho^{-1}$ dependence breaks down and sources can become hard to detect at all.
Indeed, at $\rho = 8$, the maximum residual and error of Eq.~\ref{eq:analytic_emin} is 0.015 and $14\%$ respectively, whereas for all $\rho \geq 9$ this improves to 0.0055 and $11\%$ respectively.
Outside of the tested frequency range the \LISA sensitivity is expected to decrease quite rapidly.

The minimum detectable eccentricities in Fig.~\ref{fig:min_ecc}, $e_{\rm min}\sim 0.1$, are fairly large in comparison to other types of sources in \LISA.
The difficulty in measuring eccentricity is a consequence of the small amount of information contained in the quasi-monochromatic GW signals. 
This contrasts strongly with \LISA's ability to measure much smaller eccentricities in other, broadband signals.
For example, when observing supermassive binary black hole mergers, eccentricities as small as $e\lesssim 10^{-2}$ are expected to be measurable \citep[see, for example,][]{2023arXiv230713367G}.
Or, for chirping stellar mass black holes near the upper end of the \LISA sensitive band, eccentricities as small as $e\lesssim 10^{-3}$ are expected to be measurable \citep[see, for example,][]{2022arXiv220403423K}.
Or, for extreme-mass-ratio inspirals, eccentricities as small as $e\lesssim 10^{-4}$ are expected to be measurable \citep[see, for example,][]{2004PhRvD..69h2005B,Babak_EMRI_2017}.

The optimum frequency where \LISA's sensitivity to eccentricity in quasi-monochromatic binaries is best is around $\sim 2 \,\mathrm{mHz}$. 
This is because this places the third harmonic (which is usually the second loudest) at a frequency of $\sim 1.5\times 2=3\,\mathrm{mHz}$ near the minimum of the \LISA noise bucket (see Fig.~\ref{fig:freq:ecc}).
Measuring the first subdominant harmonic (usually $n=3$) confirms that the source is eccentric.

The actual number of NS+WD systems detectable by \LISA and the fraction of those that will have a measurable eccentricity by this criterion can be assessed using binary population synthesis. 
This has been done by \citet{2024MNRAS.530..844K} where it is shown that (using their fiducial model) there are 105 \LISA-detectable NS+WD binaries of which 25 have a eccentricity that is measurable by \LISA (as assessed using Eq.~\ref{eq:analytic_emin}). 
These numbers are rather uncertain; the range of population models considered give a range 75-357 for the total number of detectable NS+WD systems and 6-68 for those with measurable eccentricities as assessed using Eq.~\ref{eq:analytic_emin} \citep[][Table 1]{2024MNRAS.530..844K}.

\subsection{Determination of the Component Masses}\label{sec:comp_masses}

It will generally not be possible to determine the component masses of quasi-monochromatic binaries observed by \LISA from their GW signals alone. 
Even if the binary has a detectable chirp, $\dot{f}_{\rm GW}$, that we are confident is due solely to the emission of GWs, only the chirp mass combination $\mathcal{M}_c$ can be determined (see Eq.~\ref{eq:chirp}).

High-eccentricity quasi-monochromatic binaries are an exception to this.
This section considers the possibility of using the measured GW chirp \emph{and} the periastron precession frequency to determine the individual component masses of the binary.
This is possible because $\dot{f}_{\rm GW}$ depends on the chirp mass (see Eq.~\ref{eq:chirp}), whereas $f_{\rm PP}$ depends on the total mass (see Eq.~\ref{eq:periastron_prec_freq}), and measuring both breaks the $m_1$, $m_2$ parameter degeneracy.
Measuring the individual component masses and determining that at least one is above $\gtrsim 1.35\,M_\odot$ would be the most direct way of confirming the presence of a NS in a quasi-monochromatic binary with \LISA alone.

Here it is assumed that the binary contains two point masses interacting only gravitationally with no external or environmental effects. 
In this situation, the measured signal parameters $\{\mathcal{A}, f_{\rm GW}, \dot{f}_{\rm GW}, f_{\rm PP}\}$ can be related to the physical binary parameters $\{m_1, m_2, a, D\}$ as described in Sec.~\ref{sec:physics}.
It should be emphasised that if these assumptions fail to hold, biased parameter estimates will be obtained.

There are several possible cases to be considered:
\begin{enumerate}[leftmargin=0.5cm, labelwidth=0.4cm]
    \item If the binary is exactly monochromatic and only $f_{\rm GW}$ is measured, then only the Kepler parameter combination $M/a^3$ can be inferred (Eq.~\ref{eq:KeplerIII}). This reveals little about the nature of the binary.
    \item If $\dot{f}_{\rm GW}$ can also be measured, then the chirp mass $\mathcal{M}_c$ can be inferred (Eq.~\ref{eq:chirp}); see \cite{2021MNRAS.502.5576K}. However, knowing $\mathcal{M}_c$ alone doesn't unambiguously determine if the binary contains a NS. 
    \item If $e$ and $f_{\rm PP}$ (but not $\dot{f}_{\rm GW}$) can also be measured, then the total mass $M$ can be inferred (Eq.~\ref{eq:periastron_prec_freq}); see \cite{2001PhRvL..87y1101S}. 
    The eccentric orbit suggests the binary evolved through a supernova, which is suggestive of a NS.
    However, knowing $M$ alone doesn't unambiguously determine if the binary contains a NS.
    \item If $f_{\rm GW}$, $\dot{f}_{\rm GW}$, $e$ and $f_{\rm PP}$ can all be measured, then it is possible to infer the orbital semi-major axis $a$ and the individual component masses $m_1$ and $m_2$ separately. 
    Together with the fact that the binary is eccentric, this is the strongest possible evidence from GWs alone that the binary contains a NS.
\end{enumerate}

Fig.~\ref{fig:golden_binary} shows parameter estimation results for a binary in case (iv).
Because not all combinations of the signal parameters correspond to a physical, non-interacting binary; when converting to the binary parameters, additional prior cuts have to be imposed (see discussion in Sec.~\ref{sec:physics}). 
This is achieved in practice by discarding posterior samples with nonphysical parameters combinations; for the $e=0.1, 0.05, 0.01$ binaries in Fig.~\ref{fig:golden_binary}, the fractions of discarded samples were 0.28, 0.43 and 0.47 respectively.
The posteriors on the physical parameters in Fig.~\ref{fig:golden_binary} were plotted with the remaining physically valid samples.
If, in a real analysis, the majority of samples are nonphysical, this may indicate that the source is not an isolated, non-interacting binary.

The top panel of Fig.~\ref{fig:golden_binary} is designed to show how the simultaneous measurements of the chirp mass (from the measurement of $\dot{f}_{\rm}$) and the total mass (from the measurement of $f_{\rm PP}$) break the mass degeneracy.
The primary mass is determined to be greater than $1.3\,M_\odot$ with 90\% confidence; strongly suggesting that this is a NS.

\begin{figure*}
    \centering
    {\includegraphics[width=0.98\textwidth]{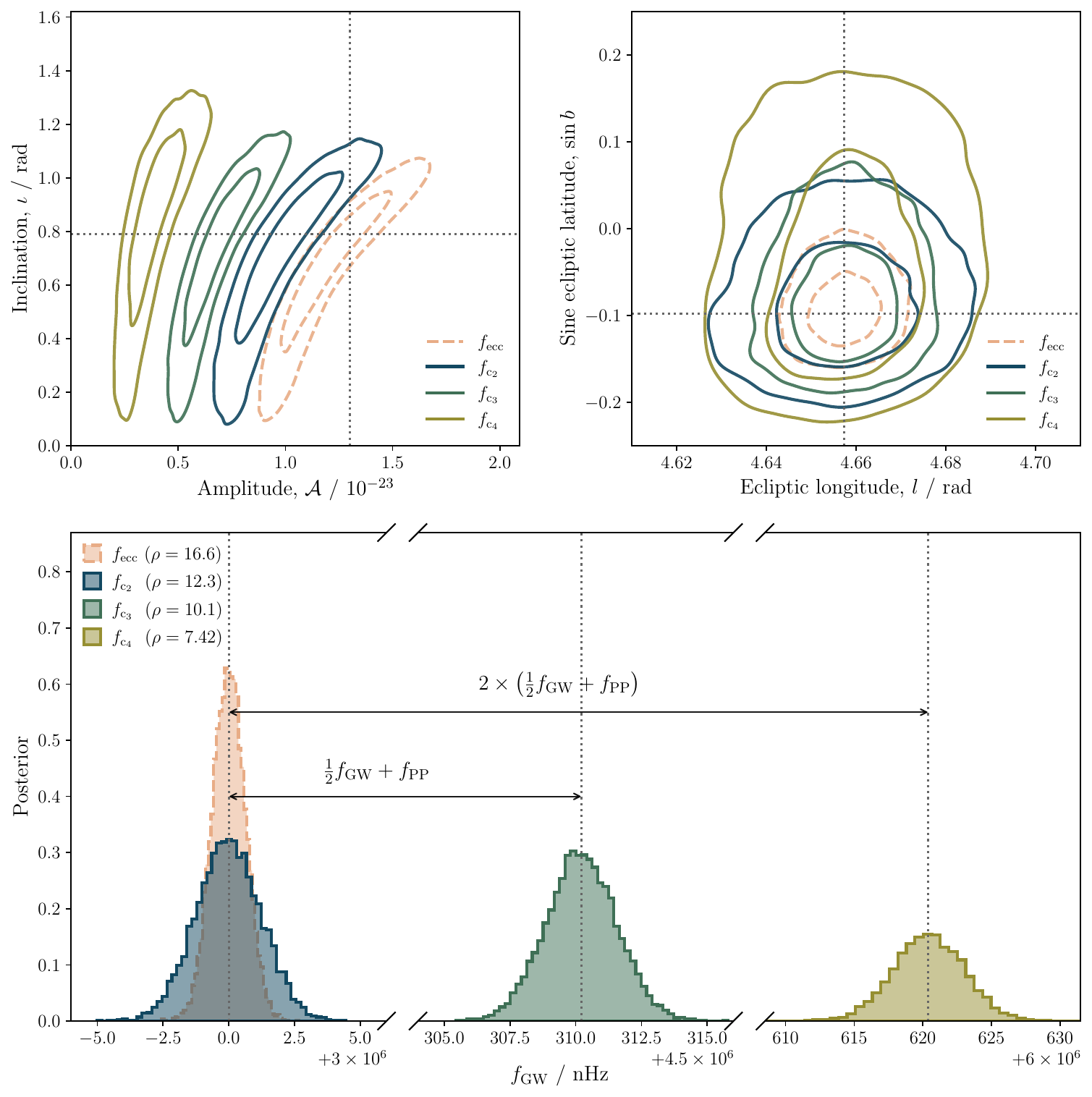}}
    \caption{ \label{fig:circ_ecc_PE}
        Posteriors illustrating the possibility of confusing eccentric harmonics for separate, circular sources.
        The eccentric source used for the injection had parameters $f_{\rm GW, ecc}=3.0\,\mathrm{mHz}$, $f_{\rm PP}= 310\,\mathrm{nHz}$, $\mathcal{A}=1.3\times 10^{-23}$, $\iota=\pi/4$, $e=0.3$ and was in the direction of the Galactic centre. 
        The total SNR of the injection was $\rho=21.7$. 
        The individual harmonics of this eccentric injection can be mistaken for separate, circular binaries.
        Results are shown for four independent parameter estimation analyses: recovery with an eccentric model (ecc, the truth), recovery with a circular model focusing on frequencies around $\sim 3.0\,\mathrm{mHz}$ ($\mathrm{c}_2$, the $n=2$ harmonic of the injected source), recovery with a circular model focusing on frequencies around $\sim 4.5\,\mathrm{mHz}$ ($\mathrm{c}_3$), and recovery with a circular model focusing on frequencies around $\sim 6\,\mathrm{mHz}$ ($\mathrm{c}_4$). 
        Additional circular analysis were performed targeting the $n=1$ and $n=5$ harmonics; however, nothing was recovered because these harmonics were too quiet.
        In all 2-dimensional plots, pairs of iso-probability contours enclose 50\% and 90\% of the posterior probability. 
        We emphasise that the results in this figure were obtained with zero-noise injections. 
        This was done for simplicity. 
        If realistic noise were to be included the effect would be to shift the posteriors of the frequency for each the individual circular binaries up or down by an amount consistent with the posterior width. 
        This will make the process of identifying these circular sources as harmonics of a single eccentric source slightly harder than it appears here; a detailed analysis of this problem involving a large number of noisy injections is beyond the scope of this paper.
    }
\end{figure*}

Fig.~\ref{fig:golden_binary} also includes vertical lines separating regions of the primary mass range where we expect either WDs or NSs. These boundaries are determined using the fiducial NS+WD mock population in \cite{2024MNRAS.530..844K} which was assembled using binary population synthesis. 
In this model the WD component mass probability density peaks $\sim 0.9\,M_\odot$ with a tail extending up to the Chandrasekhar limit ($\sim1.4\,M_\odot$), although such high masses are significantly less likely and the vast majority of primary objects are below $1\,M_\odot$. 
The minimum NS mass in NS+WD binaries is $\sim 1.22\,M_\odot$. 
Comparing these boundaries to observations of WD+WD systems poses challenges, particularly in constraining the mass of the primary star (the more massive of the pair). 
This task can be complicated by the secondary star, which is often brighter and, consequently, presents difficulties in characterising the primary's properties. 
Nevertheless, we can compare these boundaries to the mass distribution of single WDs; in spectroscopic samples, WDs with masses exceeding $1\,M_\odot$ are highly improbable and likely to be a result of a merger \citep[][]{kep15,2020A&A...636A..31T,2023MNRAS.518.2341K}. The distribution of NS masses in binaries has a mean at $\sim1.28\,M_\odot$ with a dispersion of $0.24\,M_\odot$ \citep[][see also \citealt{2019ApJ...876...18F}]{2012ApJ...757...55O}. 

For the other source parameters not shown in Fig.~\ref{fig:golden_binary}, orbital eccentricity has only a minor impact on our ability to measure them compared to the case of circular quasi-monochromatic binary. 
The parameters that show the largest effect are the sky position.
For the binary in Fig.~\ref{fig:golden_binary} with $e=0.1$, \LISA can localise the source to a 90\% sky area of $\Omega_{90}=7.9\,\mathrm{deg}^2$ whereas for the same binary with $e=0.01$ this area increases to $\Omega_{90}=16.5\,\mathrm{deg}^2$. 
Much of this difference can be attributed to the increased SNR of the eccentric source (34.8 compared to 27.2), the expected scaling being $\Omega_{90}\propto\rho^{-2}$. 
The remainder of the difference is likely due to the presence of some signal power at higher frequencies for the eccentric binary, where \LISA's angular resolution is known to be better \citep{1998PhRvD..57.7089C}.

\subsection{Misidentification of Harmonics as Circular Sources}\label{sec:results_confusion}

Each harmonic of an eccentric quasi-monochromatic binary appears in the \LISA data as a narrow-band feature in the frequency spectrum.
It is possible (in fact, very easy) to mistake these features as being produced by a number of separate circular quasi-monochromatic binaries, where the number of such circular sources is simply the number of harmonics with an SNR above the detection threshold (approximately $\rho\geq 6$).
This sort of source confusion has implications for the design of the \LISA global fit (see Sec.~\ref{sec:global fit}).

Using the waveforms described in Sec.~\ref{sec:physics}, an eccentric source that generates $m$ harmonics above the SNR threshold can be (mis)modelled perfectly by $m$ circular sources, meaning the overlap between the circular sources and the eccentric harmonics is unity.
The (fictitious) circular templates that give this perfect overlap all have the same sky position and inclination as the (true) eccentric source.
The initial frequency of the circular template matching the $n^{\rm th}$ harmonic of the eccentric source is given by
\begin{align} \label{eq:circ_ecc_f}
    f_{\mathrm{GW}, \mathrm{c}_n} = \frac{n}{2}f_{\rm GW, ecc} + (n-2)f_{\rm PP}.
\end{align}
Similarly, the chirp rate  of the circular template matching the $n^{\rm th}$ harmonic of the eccentric source is given by
\begin{align} \label{eq:circ_ecc_fdot}
    \dot{f}_{\mathrm{GW}, \mathrm{c}_n} = \frac{n}{2}\dot{f}_{\rm GW, ecc}.
\end{align}
Finally, the amplitude of the circular template matching the $n^{\rm th}$ harmonic of the eccentric source is given by Eqs.~\ref{eq:amp_analytical} or \ref{eq:circ_ecc_A}.
In general, the $n=2$ harmonic has the largest amplitude (at least for moderate eccentricities, $e\lesssim 0.29$) and this is a decreasing function of eccentricity, while the amplitudes of the other modes are smaller and are increasing functions of the eccentricity. 
This behaviour can also be seen clearly in Fig.~\ref{fig:freq:ecc}.

In order to investigate the possibility of confusing eccentric harmonics for circular sources another, zero-noise injection was performed using a loud eccentric source with $e=0.3$ and $\rho=21.7$.
Parameter estimation was first performed on this signal using the (correct) eccentric waveform model. 
Then, parameter estimation was performed again using an (incorrect) circular waveform model targeting only the data in a narrow frequency range around each harmonic.
A subset of the parameter estimation results are shown in Fig.~\ref{fig:circ_ecc_PE}.
For this system, the SNRs in the first five harmonics were $\rho_{n=1}=0.273$, $\rho_{n=2}=14.6$, $\rho_{n=3}=13.7$, $\rho_{n=4}=7.42$, and $\rho_{n=5}=3.44$. 
It was found that the circular analyses could only recover the $n=2$, $3$, and $4$ harmonics, since these are the harmonics above the detectability threshold of $\sim 6$. 

As expected, the eccentric analyses correctly recover all the injected source parameters. 
The circular analyses recover parameter values that can be related back to the parameters of the eccentric injection.
For example, the frequency (bottom panel) of the $n^{\rm th}$ harmonic is a function of the frequency of the eccentric injection, $f_{\rm GW, ecc}$, its periastron precession frequency, $f_{\rm PP}$, and the harmonic index, $n$ (see Eq.~\ref{eq:circ_ecc_f}).
The sky positions (top right) of all the (fictitious) circular sources are consistent with that of the eccentric injection but with larger uncertainties corresponding to the reduced SNR when only a single harmonic is being analysed.
Likewise, the inclinations (top left) of the circular sources are consistent with the eccentric injection but with larger uncertainties. 
The amplitudes (top left) of the circular sources are lower than that of the eccentric injection, again corresponding to the reduced SNR in a single harmonic and given by Eq.~\ref{eq:amp_analytical}. 

In a global fit scenario, the simpler (and more common) circular sources will likely be targeted first, and so this misidentification of eccentric harmonics as separate circular sources is likely to occur (this is discussed further in Sec.~\ref{sec:global fit}).

\subsection{Reclassification of Circular Sources as an Eccentric Source}\label{subsec:reclassify}

\begin{figure}
    \centering
    \includegraphics[width=0.98\columnwidth]{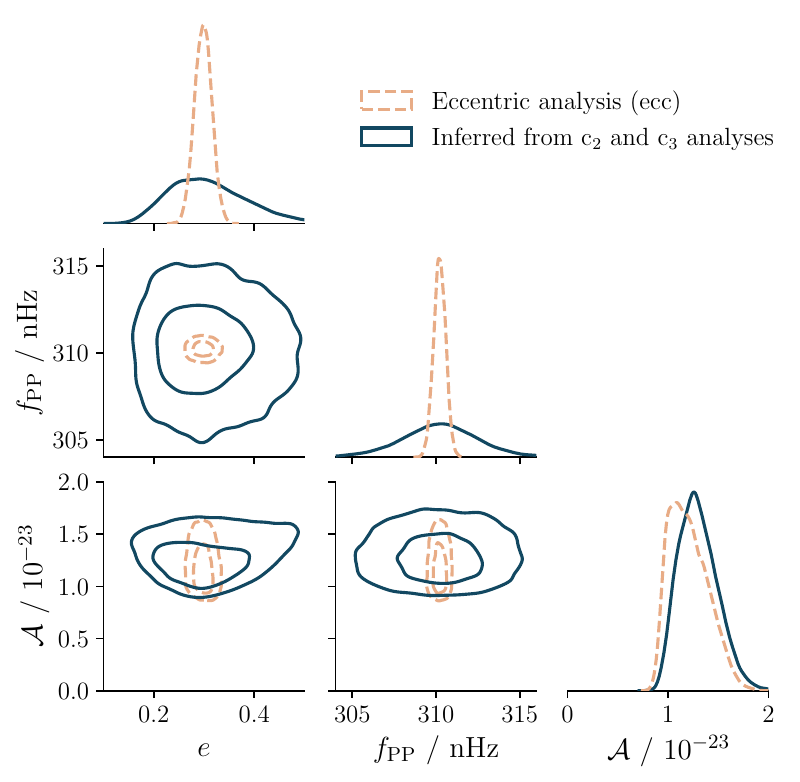}
    \caption{
        Posteriors on selected parameters for the same eccentric source analysed in Fig.~\ref{fig:circ_ecc_PE}.
        The pale dashed line shows the results from the eccentric analysis (using the same waveform model for the parameter estimation analysis as was used for the injection). 
        The darker solid line shows the rough results that can be inferred by combining the results of two circular parameter estimation analyses targeting the $n=2$ and $n=3$ harmonics (the $\mathrm{c}_2$ and $\mathrm{c}_3$ analyses described in Fig.~\ref{fig:circ_ecc_PE}). 
        The three parameters shown are the eccentricity $e$, the periastron precession frequency $f_{\rm PP}$ and the strain amplitude, $\mathcal{A}$.
        The spacing of the recovered frequencies from the two circular analyses (the fact that the frequency of the $\mathrm{c}_3$ source is not exactly 3/2 times the $\mathrm{c}_2$ source) encodes $f_{\rm PP}$ and the ratio of the recovered amplitudes from the two circular analyses encodes $e$.
        In all 2-dimensional plots, pairs of iso-probability contours enclose 50\% and 90\% of the posterior probability.
    }
    \label{fig:ecc_par_from_circ}
\end{figure}

\setlength{\tabcolsep}{1.7pt}
\begin{table}
	\centering
	\caption{ \label{tab:D}
        The individual harmonic SNRs for the source analysed in  Fig.~\ref{fig:circ_ecc_PE} and Secs.~\ref{sec:results_confusion} and \ref{subsec:reclassify}. 
        The total SNR (Eq.~\ref{eq:snr}) is $\rho = 21.67$.
        Even for this source with a relatively large eccentricity of $e=0.3$, just three harmonics ($n=2$, 3 and 4; highlighted) contain 97\% of the total squared SNR.
    }
	\begin{tabular}{c|cccccccccccc} 
		\hline
        $n$ & 1 & \cellcolor{blue!15} 2 & \cellcolor{blue!14} 3 & \cellcolor{blue!8} 4 & \cellcolor{blue!4} 5 & \cellcolor{blue!2} 6 & \cellcolor{blue!1} 7 & 8 & 9 & 10 & 11 & 12 \\
        $\rho_n$ & 0.27 & \cellcolor{blue!15} 14.58 & \cellcolor{blue!14} 13.69 & \cellcolor{blue!8} 7.42 &  \cellcolor{blue!4} 3.44 & \cellcolor{blue!2} 1.51 & \cellcolor{blue!1} 0.65 & 0.27 & 0.11 & 0.04 & 0.03 & 0.01 \\
        \hline
	\end{tabular}
\end{table}

As mentioned in the previous section, the first stage of a global fit is likely to involve searching for all circular sources that can be found in the data, as these are the most numerous and simplest \LISA sources. 
In this scenario, a search for eccentric sources will only be performed at a later stage.
As part of this search, it will be possible to identify pairs of circular source candidates that are likely two harmonics of a single, eccentric source (e.g.\ they have overlapping sky and inclination posteriors and with frequencies that are closely related by a factor of $3/2$). 
Using the existing circular parameter estimates for this pair of sources, it is possible to get a crude estimate for the eccentric source parameters without any additional parameter estimation; the frequency spacing gives you the periastron precession frequency, and the amplitudes give you the eccentricity.
This can be used to verify if two circular candidates are in fact harmonics of a single eccentric source (by seeing if the inferred $f_\mathrm{PP}$ and $e$ are astrophysically reasonable, for example).
In a global fit the multiple circular waveforms will want to be swapped out for a single eccentric waveform, and this procedure could also be used to efficiently initialise the parameters of the eccentric waveform.
This section demonstrates the idea of estimating eccentric parameters from the circular parameter estimation for the eccentric source in Fig.~\ref{fig:circ_ecc_PE}. 
For simplicity we use only the circular analyses of the $n=2$ and $n=3$ harmonics.

Let $f_{\rm GW, c_n}$ denote the value of the frequency of a circular template being used to analyse the $n^{\rm th}$ harmonic of an eccentric source. 
The relationship between the circular frequencies and the properties of the eccentric source is given in Eq.~\ref{eq:circ_ecc_f}.
This was used with $n=2$ and $n=3$ to infer the value of $f_{\rm PP}$ for the eccentric source.
The strain amplitudes $\mathcal{A}_{\mathrm{c}_n}$ of the circular templates can be related to the eccentricity.
The expressions for the individual amplitudes of the $n=2$ and $n=3$ harmonics are given by Eq.~\ref{eq:amp_analytical}.
These expressions were inverted numerically to find $\mathcal{A}$ and $e$ as functions of $\mathcal{A}_{\mathrm{c}_2}$ and $\mathcal{A}_{\mathrm{c}_3}$. 
The resulting posteriors are shown in Fig.~\ref{fig:ecc_par_from_circ}.

As can be seen from Fig.~\ref{fig:ecc_par_from_circ}, the posteriors on the eccentric parameters inferred from the circular analyses are broader (i.e.\ less informative) than those from the eccentric analysis.
This is expected because the circular analyses are not analysing all of the available signal coherently, and this results in a loss of SNR. 
The SNRs in the individual harmonics for this source are given in Table \ref{tab:D}.

\section{Discussion and implications for the design of the \LISA global fit}\label{sec:global fit}

A few percent of the quasi-monochromatic binaries that \LISA will observe may be eccentric. 
A non-zero eccentricity suggests that the source may contain a NS and merit followup analyses.
If both an increasing frequency and advancing periastron can be measured in a non-interacting binary then the individual component masses can be determined, helping to confirm the presence of a NS. 
This has been demonstrated for several simulated sources and \LISA's ability to measure eccentricity has been quantified across parameter space.
Using the results for the \LISA minimum measurable eccentricity in combination with binary population synthesis, \cite{2024MNRAS.530..844K} expect between 6-68 of the 75-357 detectable WD+NS \LISA sources to have a measurable eccentricity.

The widely-separated frequency harmonics of eccentric quasi-monochromatic binaries are unusual among GW sources, which generally consist of overlapping broadband modes. 
A multi-harmonic heterodyning approach that allows all these harmonics to be efficiently analysed coherently has been used as part of our Bayesian parameter inference. 
The benefits of analysing all harmonics together (including harmonics that contain a small fraction of the total SNR), as opposed analysing harmonics individually as if they were circular sources has been demonstrated (see Appendix \ref{app:varyN}).

Ultimately, both eccentric and circular sources will have to be analysed together as part of the \LISA \emph{global fit}.
This consists of the simultaneous detection and characterisation of the numerous GW sources, instrumental features, glitches, and noise sources in the \LISA data.
The details of this complex process are yet to be finalised \citep[for an early prototype showing what this might look like, see][]{2023PhRvD.107f3004L}. 
However, it is clear that a hierarchical and iterative approach will be necessary, with the simplest sources being provisionally identified and analysed first before the rest are tackled. 
This process will iterated many times before it eventually converges. 

The first stage of the global fit will almost certainly target the easily-resolvable circular (i.e.\ single-harmonic) Galactic binaries as these are both the simplest and most numerous \LISA source type.
As has been shown here, this first stage may incorrectly identify a harmonic of a loud eccentric source as a circular source. 
(If the eccentric source is sufficiently loud and eccentric, then several harmonics might be incorrectly identified as separate circular sources.)
It will be necessary for the later stages of the global fit to correct these misidentifications, either by looking for circular source candidates in the still-evolving catalogue that are likely harmonics of a single source (e.g., with frequencies that differ by a factor of approximately $n/2$ and that have consistent sky positions, frequency derivatives and inclination angles) and removing/replacing them, or by regularly performing eccentric versus circular model comparison on every candidate in the continually-evolving catalogue.
This highlights a general feature required by any \LISA global fit; while the algorithm as a whole is still running, the inputs and outputs of its different components are only provisional and must be subject to revision and modification by other parts of the algorithm. 

It is also interesting to note that, what with the large number of circular sources in the \LISA band, it is possible that multiple (actually circular) sources could by chance have a frequency spacing that makes them appear as a false-eccentric source.
Multiple sources with overlapping sky and inclination posteriors is not surprising, since the uncertainty on these is large compared to the separation of sources on the sky.
Frequency is the best measured and will be the best distinguisher. However, note that there is some freedom in the spacing introduced by the periastron precession factor, which will not be known beforehand.
The combined phase-polarization measurement could also be useful, but only if more than two modes are detected (the phase-polarization combination for each harmonic will be regularly spaced, but that spacing is not known beforehand so this is not useful when only two harmonics are measured).
And note that this will be a bigger problem earlier in the mission, when less SNR has accumulated and posteriors are wider. 

Eccentric quasi-monochromatic binaries are a possible \LISA source type that has hitherto received little attention in the literature. 
While, like circular Galactic binaries, these sources are relatively simple to model they nevertheless pose a new analysis challenge due to the necessity of analysing multiple, widely separated frequency harmonics and the possibility of confusion with the more numerous circular sources.
This paper has shown how these problems may be tackled and how, by identifying eccentric sources, \LISA may be used as a new tool for the discovery of NS+WD Galactic binaries.

\section*{Acknowledgements}
 
We thank the developers of the \textsc{Balrog} codesuite, including those who are not authors here.
Computational resources used for this work were provided by the University of Birmingham’s BlueBEAR High Performance Computing facility.
CJM and AK acknowledge the support of the UK Space Agency grant, no.\ ST/V002813/1.
NP and DR acknowledge support provided by Research England under the Enhancing Research Culture funding stream, grant ERC-3: \emph{The UoB Astronomy Summer School}.

\section*{Data Availability}
 
A data release with posterior samples and notebooks to reproduce figures is made publicly available on Zenodo \citep{moore_christopher_j_2023_8417770}.


\clearpage
\bibliographystyle{mnras}
\bibliography{refs}

\begin{thebibliography}{}
\makeatletter
\relax
\def\mn@urlcharsother{\let\do\@makeother \do\$\do\&\do\#\do\^\do\_\do\%\do\~}
\def\mn@doi{\begingroup\mn@urlcharsother \@ifnextchar [ {\mn@doi@}
  {\mn@doi@[]}}
\def\mn@doi@[#1]#2{\def\@tempa{#1}\ifx\@tempa\@empty \href
  {http://dx.doi.org/#2} {doi:#2}\else \href {http://dx.doi.org/#2} {#1}\fi
  \endgroup}
\def\mn@eprint#1#2{\mn@eprint@#1:#2::\@nil}
\def\mn@eprint@arXiv#1{\href {http://arxiv.org/abs/#1} {{\tt arXiv:#1}}}
\def\mn@eprint@dblp#1{\href {http://dblp.uni-trier.de/rec/bibtex/#1.xml}
  {dblp:#1}}
\def\mn@eprint@#1:#2:#3:#4\@nil{\def\@tempa {#1}\def\@tempb {#2}\def\@tempc
  {#3}\ifx \@tempc \@empty \let \@tempc \@tempb \let \@tempb \@tempa \fi \ifx
  \@tempb \@empty \def\@tempb {arXiv}\fi \@ifundefined
  {mn@eprint@\@tempb}{\@tempb:\@tempc}{\expandafter \expandafter \csname
  mn@eprint@\@tempb\endcsname \expandafter{\@tempc}}}

\bibitem[\protect\citeauthoryear{{Amaro-Seoane} et~al.,}{{Amaro-Seoane}
  et~al.}{2017}]{2017arXiv170200786A}
{Amaro-Seoane} P.,  et~al., 2017, \mn@doi [arXiv e-prints]
  {10.48550/arXiv.1702.00786}, \href
  {https://ui.adsabs.harvard.edu/abs/2017arXiv170200786A} {p. arXiv:1702.00786}

\bibitem[\protect\citeauthoryear{{Babak} et~al.,}{{Babak}
  et~al.}{2017}]{Babak_EMRI_2017}
{Babak} S.,  et~al., 2017, \mn@doi [\prd] {10.1103/PhysRevD.95.103012}, 95,
  103012

\bibitem[\protect\citeauthoryear{{Babak}, {Hewitson}  \& {Petiteau}}{{Babak}
  et~al.}{2021}]{2021_Babak_PSD}
{Babak} S.,  {Hewitson} M.,   {Petiteau} A.,  2021, arXiv e-prints

\bibitem[\protect\citeauthoryear{{Bandopadhyay} \& {Moore}}{{Bandopadhyay} \&
  {Moore}}{2023}]{2023arXiv230518048B}
{Bandopadhyay} D.,  {Moore} C.~J.,  2023, \mn@doi [arXiv e-prints]
  {10.48550/arXiv.2305.18048}, \href
  {https://ui.adsabs.harvard.edu/abs/2023arXiv230518048B} {p. arXiv:2305.18048}

\bibitem[\protect\citeauthoryear{{Barack} \& {Cutler}}{{Barack} \&
  {Cutler}}{2004}]{2004PhRvD..69h2005B}
{Barack} L.,  {Cutler} C.,  2004, \mn@doi [\prd] {10.1103/PhysRevD.69.082005},
  \href {https://ui.adsabs.harvard.edu/abs/2004PhRvD..69h2005B} {69, 082005}

\bibitem[\protect\citeauthoryear{{Bender} \& {Hils}}{{Bender} \&
  {Hils}}{1997}]{1997CQGra..14.1439B}
{Bender} P.~L.,  {Hils} D.,  1997, \mn@doi [Classical and Quantum Gravity]
  {10.1088/0264-9381/14/6/008}, \href
  {https://ui.adsabs.harvard.edu/abs/1997CQGra..14.1439B} {14, 1439}

\bibitem[\protect\citeauthoryear{{Bobrick}, {Davies}  \& {Church}}{{Bobrick}
  et~al.}{2017}]{2017MNRAS.467.3556B}
{Bobrick} A.,  {Davies} M.~B.,   {Church} R.~P.,  2017, \mn@doi [\mnras]
  {10.1093/mnras/stx312}, \href
  {https://ui.adsabs.harvard.edu/abs/2017MNRAS.467.3556B} {467, 3556}

\bibitem[\protect\citeauthoryear{{Breivik} et~al.,}{{Breivik}
  et~al.}{2020}]{2020ApJ...898...71B}
{Breivik} K.,  et~al., 2020, \mn@doi [\apj] {10.3847/1538-4357/ab9d85}, \href
  {https://ui.adsabs.harvard.edu/abs/2020ApJ...898...71B} {898, 71}

\bibitem[\protect\citeauthoryear{{Brown}, {Lee}, {Portegies Zwart}  \&
  {Bethe}}{{Brown} et~al.}{2001}]{2001ApJ...547..345B}
{Brown} G.~E.,  {Lee} C.~H.,  {Portegies Zwart} S.~F.,   {Bethe} H.~A.,  2001,
  \mn@doi [\apj] {10.1086/318327}, \href
  {https://ui.adsabs.harvard.edu/abs/2001ApJ...547..345B} {547, 345}

\bibitem[\protect\citeauthoryear{{Buscicchio}, {Roebber}, {Goldstein}  \&
  {Moore}}{{Buscicchio} et~al.}{2019}]{2019PhRvD.100h4041B}
{Buscicchio} R.,  {Roebber} E.,  {Goldstein} J.~M.,   {Moore} C.~J.,  2019,
  \mn@doi [\prd] {10.1103/PhysRevD.100.084041}, \href
  {https://ui.adsabs.harvard.edu/abs/2019PhRvD.100h4041B} {100, 084041}

\bibitem[\protect\citeauthoryear{{Buscicchio}, {Klein}, {Roebber}, {Moore},
  {Gerosa}, {Finch}  \& {Vecchio}}{{Buscicchio}
  et~al.}{2021}]{2021PhRvD.104d4065B}
{Buscicchio} R.,  {Klein} A.,  {Roebber} E.,  {Moore} C.~J.,  {Gerosa} D.,
  {Finch} E.,   {Vecchio} A.,  2021, \mn@doi [\prd]
  {10.1103/PhysRevD.104.044065}, \href
  {https://ui.adsabs.harvard.edu/abs/2021PhRvD.104d4065B} {104, 044065}

\bibitem[\protect\citeauthoryear{{Cutler}}{{Cutler}}{1998}]{1998PhRvD..57.7089C}
{Cutler} C.,  1998, \mn@doi [\prd] {10.1103/PhysRevD.57.7089}, \href
  {https://ui.adsabs.harvard.edu/abs/1998PhRvD..57.7089C} {57, 7089}

\bibitem[\protect\citeauthoryear{Damour \& Deruelle}{Damour \&
  Deruelle}{1985}]{AIHPA_1985__43_1_107_0}
Damour T.,  Deruelle N.,  1985, Annales de l'I.H.P. Physique th\'eorique, 43,
  107

\bibitem[\protect\citeauthoryear{{Damour}, {Iyer}  \& {Sathyaprakash}}{{Damour}
  et~al.}{2001}]{2001PhRvD..63d4023D}
{Damour} T.,  {Iyer} B.~R.,   {Sathyaprakash} B.~S.,  2001, \mn@doi [\prd]
  {10.1103/PhysRevD.63.044023}, \href
  {https://ui.adsabs.harvard.edu/abs/2001PhRvD..63d4023D} {63, 044023}

\bibitem[\protect\citeauthoryear{{Damour}, {Iyer}  \& {Sathyaprakash}}{{Damour}
  et~al.}{2005}]{2005PhRvD..72b9902D}
{Damour} T.,  {Iyer} B.~R.,   {Sathyaprakash} B.~S.,  2005, \mn@doi [\prd]
  {10.1103/PhysRevD.72.029902}, \href
  {https://ui.adsabs.harvard.edu/abs/2005PhRvD..72b9902D} {72, 029902}

\bibitem[\protect\citeauthoryear{{Davies}, {Ritter}  \& {King}}{{Davies}
  et~al.}{2002}]{2002MNRAS.335..369D}
{Davies} M.~B.,  {Ritter} H.,   {King} A.,  2002, \mn@doi [\mnras]
  {10.1046/j.1365-8711.2002.05594.x}, \href
  {https://ui.adsabs.harvard.edu/abs/2002MNRAS.335..369D} {335, 369}

\bibitem[\protect\citeauthoryear{{Farmer} \& {Phinney}}{{Farmer} \&
  {Phinney}}{2003}]{2003MNRAS.346.1197F}
{Farmer} A.~J.,  {Phinney} E.~S.,  2003, \mn@doi [\mnras]
  {10.1111/j.1365-2966.2003.07176.x}, \href
  {https://ui.adsabs.harvard.edu/abs/2003MNRAS.346.1197F} {346, 1197}

\bibitem[\protect\citeauthoryear{{Farrow}, {Zhu}  \& {Thrane}}{{Farrow}
  et~al.}{2019}]{2019ApJ...876...18F}
{Farrow} N.,  {Zhu} X.-J.,   {Thrane} E.,  2019, \mn@doi [\apj]
  {10.3847/1538-4357/ab12e3}, \href
  {https://ui.adsabs.harvard.edu/abs/2019ApJ...876...18F} {876, 18}

\bibitem[\protect\citeauthoryear{{Finch} et~al.,}{{Finch}
  et~al.}{2023}]{2023MNRAS.522.5358F}
{Finch} E.,  et~al., 2023, \mn@doi [\mnras] {10.1093/mnras/stad1288}, \href
  {https://ui.adsabs.harvard.edu/abs/2023MNRAS.522.5358F} {522, 5358}

\bibitem[\protect\citeauthoryear{{Garg}, {Tiwari}, {Derdzinski}, {Baker},
  {Marsat}  \& {Mayer}}{{Garg} et~al.}{2023}]{2023arXiv230713367G}
{Garg} M.,  {Tiwari} S.,  {Derdzinski} A.,  {Baker} J.,  {Marsat} S.,   {Mayer}
  L.,  2023, \mn@doi [arXiv e-prints] {10.48550/arXiv.2307.13367}, \href
  {https://ui.adsabs.harvard.edu/abs/2023arXiv230713367G} {p. arXiv:2307.13367}

\bibitem[\protect\citeauthoryear{{Georgousi}, {Karnesis}, {Korol}, {Pieroni}
  \& {Stergioulas}}{{Georgousi} et~al.}{2023}]{2023MNRAS.519.2552G}
{Georgousi} M.,  {Karnesis} N.,  {Korol} V.,  {Pieroni} M.,   {Stergioulas} N.,
   2023, \mn@doi [\mnras] {10.1093/mnras/stac3686}, \href
  {https://ui.adsabs.harvard.edu/abs/2023MNRAS.519.2552G} {519, 2552}

\bibitem[\protect\citeauthoryear{{Hils}, {Bender}  \& {Webbink}}{{Hils}
  et~al.}{1990}]{1990ApJ...360...75H}
{Hils} D.,  {Bender} P.~L.,   {Webbink} R.~F.,  1990, \mn@doi [\apj]
  {10.1086/169098}, \href
  {https://ui.adsabs.harvard.edu/abs/1990ApJ...360...75H} {360, 75}

\bibitem[\protect\citeauthoryear{{Kaspi} et~al.,}{{Kaspi}
  et~al.}{2000}]{2000ApJ...543..321K}
{Kaspi} V.~M.,  et~al., 2000, \mn@doi [\apj] {10.1086/317103}, \href
  {https://ui.adsabs.harvard.edu/abs/2000ApJ...543..321K} {543, 321}

\bibitem[\protect\citeauthoryear{{Kepler} et~al.,}{{Kepler}
  et~al.}{2015}]{kep15}
{Kepler} S.~O.,  et~al., 2015, \mn@doi [\mnras] {10.1093/mnras/stu2388}, \href
  {https://ui.adsabs.harvard.edu/abs/2015MNRAS.446.4078K} {446, 4078}

\bibitem[\protect\citeauthoryear{{Kilic} et~al.,}{{Kilic}
  et~al.}{2023}]{2023MNRAS.518.2341K}
{Kilic} M.,  et~al., 2023, \mn@doi [\mnras] {10.1093/mnras/stac3182}, \href
  {https://ui.adsabs.harvard.edu/abs/2023MNRAS.518.2341K} {518, 2341}

\bibitem[\protect\citeauthoryear{{Klein}}{{Klein}}{2021}]{2021arXiv210610291K}
{Klein} A.,  2021, \mn@doi [arXiv e-prints] {10.48550/arXiv.2106.10291}, \href
  {https://ui.adsabs.harvard.edu/abs/2021arXiv210610291K} {p. arXiv:2106.10291}

\bibitem[\protect\citeauthoryear{{Klein} et~al.,}{{Klein}
  et~al.}{2022}]{2022arXiv220403423K}
{Klein} A.,  et~al., 2022, \mn@doi [arXiv e-prints]
  {10.48550/arXiv.2204.03423}, \href
  {https://ui.adsabs.harvard.edu/abs/2022arXiv220403423K} {p. arXiv:2204.03423}

\bibitem[\protect\citeauthoryear{{Korol}}{{Korol}}{2023}]{2023IAUS..363..203K}
{Korol} V.,  2023, \mn@doi [IAU Symposium] {10.1017/S1743921322000618}, \href
  {https://ui.adsabs.harvard.edu/abs/2023IAUS..363..203K} {363, 203}

\bibitem[\protect\citeauthoryear{{Korol} \& {Safarzadeh}}{{Korol} \&
  {Safarzadeh}}{2021}]{2021MNRAS.502.5576K}
{Korol} V.,  {Safarzadeh} M.,  2021, \mn@doi [\mnras] {10.1093/mnras/stab310},
  \href {https://ui.adsabs.harvard.edu/abs/2021MNRAS.502.5576K} {502, 5576}

\bibitem[\protect\citeauthoryear{{Korol} et~al.,}{{Korol}
  et~al.}{2020}]{2020A&A...638A.153K}
{Korol} V.,  et~al., 2020, \mn@doi [\aap] {10.1051/0004-6361/202037764}, \href
  {https://ui.adsabs.harvard.edu/abs/2020A&A...638A.153K} {638, A153}

\bibitem[\protect\citeauthoryear{{Korol}, {Belokurov}, {Moore}  \&
  {Toonen}}{{Korol} et~al.}{2021}]{2021MNRAS.502L..55K}
{Korol} V.,  {Belokurov} V.,  {Moore} C.~J.,   {Toonen} S.,  2021, \mn@doi
  [\mnras] {10.1093/mnrasl/slab003}, \href
  {https://ui.adsabs.harvard.edu/abs/2021MNRAS.502L..55K} {502, L55}

\bibitem[\protect\citeauthoryear{{Korol}, {Hallakoun}, {Toonen}  \&
  {Karnesis}}{{Korol} et~al.}{2022}]{2022MNRAS.511.5936K}
{Korol} V.,  {Hallakoun} N.,  {Toonen} S.,   {Karnesis} N.,  2022, \mn@doi
  [\mnras] {10.1093/mnras/stac415}, \href
  {https://ui.adsabs.harvard.edu/abs/2022MNRAS.511.5936K} {511, 5936}

\bibitem[\protect\citeauthoryear{{Korol}, {Igoshev}, {Toonen}, {Karnesis},
  {Moore}, {Finch}  \& {Klein}}{{Korol} et~al.}{2024}]{2024MNRAS.530..844K}
{Korol} V.,  {Igoshev} A.~P.,  {Toonen} S.,  {Karnesis} N.,  {Moore} C.~J.,
  {Finch} E.,   {Klein} A.,  2024, \mn@doi [\mnras] {10.1093/mnras/stae889},
  \href {https://ui.adsabs.harvard.edu/abs/2024MNRAS.530..844K} {530, 844}

\bibitem[\protect\citeauthoryear{{Kremer}, {Breivik}, {Larson}  \&
  {Kalogera}}{{Kremer} et~al.}{2017}]{2017ApJ...846...95K}
{Kremer} K.,  {Breivik} K.,  {Larson} S.~L.,   {Kalogera} V.,  2017, \mn@doi
  [\apj] {10.3847/1538-4357/aa8557}, \href
  {https://ui.adsabs.harvard.edu/abs/2017ApJ...846...95K} {846, 95}

\bibitem[\protect\citeauthoryear{{Kupfer} et~al.,}{{Kupfer}
  et~al.}{2023}]{2023arXiv230212719K}
{Kupfer} T.,  et~al., 2023, \mn@doi [arXiv e-prints]
  {10.48550/arXiv.2302.12719}, \href
  {https://ui.adsabs.harvard.edu/abs/2023arXiv230212719K} {p. arXiv:2302.12719}

\bibitem[\protect\citeauthoryear{{Li}, {Chen}, {Chen}, {Li}, {Yu}  \&
  {Han}}{{Li} et~al.}{2020}]{2020ApJ...893....2L}
{Li} Z.,  {Chen} X.,  {Chen} H.-L.,  {Li} J.,  {Yu} S.,   {Han} Z.,  2020,
  \mn@doi [\apj] {10.3847/1538-4357/ab7dc2}, \href
  {https://ui.adsabs.harvard.edu/abs/2020ApJ...893....2L} {893, 2}

\bibitem[\protect\citeauthoryear{{Lipunov}, {Postnov}  \&
  {Prokhorov}}{{Lipunov} et~al.}{1987}]{1987A&A...176L...1L}
{Lipunov} V.~M.,  {Postnov} K.~A.,   {Prokhorov} M.~E.,  1987, \aap, \href
  {https://ui.adsabs.harvard.edu/abs/1987A&A...176L...1L} {176, L1}

\bibitem[\protect\citeauthoryear{{Littenberg} \& {Cornish}}{{Littenberg} \&
  {Cornish}}{2023}]{2023PhRvD.107f3004L}
{Littenberg} T.~B.,  {Cornish} N.~J.,  2023, \mn@doi [\prd]
  {10.1103/PhysRevD.107.063004}, \href
  {https://ui.adsabs.harvard.edu/abs/2023PhRvD.107f3004L} {107, 063004}

\bibitem[\protect\citeauthoryear{{Lyne} \& {McKenna}}{{Lyne} \&
  {McKenna}}{1989}]{1989Natur.340..367L}
{Lyne} A.~G.,  {McKenna} J.,  1989, \mn@doi [\nat] {10.1038/340367a0}, \href
  {https://ui.adsabs.harvard.edu/abs/1989Natur.340..367L} {340, 367}

\bibitem[\protect\citeauthoryear{MacKay}{MacKay}{2003}]{MacKay2003}
MacKay D. J.~C.,  2003, Information Theory, Inference, and Learning Algorithms.
Copyright Cambridge University Press

\bibitem[\protect\citeauthoryear{{Mishra}, {Arun}  \& {Iyer}}{{Mishra}
  et~al.}{2015}]{2015PhRvD..91h4040M}
{Mishra} C.~K.,  {Arun} K.~G.,   {Iyer} B.~R.,  2015, \mn@doi [\prd]
  {10.1103/PhysRevD.91.084040}, \href
  {https://ui.adsabs.harvard.edu/abs/2015PhRvD..91h4040M} {91, 084040}

\bibitem[\protect\citeauthoryear{{Misner}, {Thorne}  \& {Wheeler}}{{Misner}
  et~al.}{1973}]{1973grav.book.....M}
{Misner} C.~W.,  {Thorne} K.~S.,   {Wheeler} J.~A.,  1973, {Gravitation}.
{}

\bibitem[\protect\citeauthoryear{Moore, Finch, Klein, Korol, Pham  \&
  Robins}{Moore et~al.}{2023}]{moore_christopher_j_2023_8417770}
Moore C.~J.,  Finch E.,  Klein A.,  Korol V.,  Pham N.,   Robins D.,  2023,
  {Data release for "Discovering neutron stars with LISA via measurements of
  orbital eccentricity in Galactic binaries"}, \mn@doi{10.5281/zenodo.8417769},
  \url {https://doi.org/10.5281/zenodo.8417769}

\bibitem[\protect\citeauthoryear{{Moreno-Garrido}, {Buitrago}  \&
  {Mediavilla}}{{Moreno-Garrido} et~al.}{1994}]{1994MNRAS.266...16M}
{Moreno-Garrido} C.,  {Buitrago} J.,   {Mediavilla} E.,  1994, \mn@doi [\mnras]
  {10.1093/mnras/266.1.16}, \href
  {https://ui.adsabs.harvard.edu/abs/1994MNRAS.266...16M} {266, 16}

\bibitem[\protect\citeauthoryear{{Moreno-Garrido}, {Mediavilla}  \&
  {Buitrago}}{{Moreno-Garrido} et~al.}{1995}]{1995MNRAS.274..115M}
{Moreno-Garrido} C.,  {Mediavilla} E.,   {Buitrago} J.,  1995, \mn@doi [\mnras]
  {10.1093/mnras/274.1.115}, \href
  {https://ui.adsabs.harvard.edu/abs/1995MNRAS.274..115M} {274, 115}

\bibitem[\protect\citeauthoryear{{Nelemans}, {Yungelson}, {Portegies Zwart}  \&
  {Verbunt}}{{Nelemans} et~al.}{2001a}]{2001A&A...365..491N}
{Nelemans} G.,  {Yungelson} L.~R.,  {Portegies Zwart} S.~F.,   {Verbunt} F.,
  2001a, \mn@doi [\aap] {10.1051/0004-6361:20000147}, \href
  {https://ui.adsabs.harvard.edu/abs/2001A&A...365..491N} {365, 491}

\bibitem[\protect\citeauthoryear{{Nelemans}, {Yungelson}  \& {Portegies
  Zwart}}{{Nelemans} et~al.}{2001b}]{2001A&A...375..890N}
{Nelemans} G.,  {Yungelson} L.~R.,   {Portegies Zwart} S.~F.,  2001b, \mn@doi
  [\aap] {10.1051/0004-6361:20010683}, \href
  {https://ui.adsabs.harvard.edu/abs/2001A&A...375..890N} {375, 890}

\bibitem[\protect\citeauthoryear{{Nelemans}, {Yungelson}  \& {Portegies
  Zwart}}{{Nelemans} et~al.}{2004}]{2004MNRAS.349..181N}
{Nelemans} G.,  {Yungelson} L.~R.,   {Portegies Zwart} S.~F.,  2004, \mn@doi
  [\mnras] {10.1111/j.1365-2966.2004.07479.x}, \href
  {https://ui.adsabs.harvard.edu/abs/2004MNRAS.349..181N} {349, 181}

\bibitem[\protect\citeauthoryear{{Ng} et~al.,}{{Ng}
  et~al.}{2018}]{2018MNRAS.476.4315N}
{Ng} C.,  et~al., 2018, \mn@doi [\mnras] {10.1093/mnras/sty482}, \href
  {https://ui.adsabs.harvard.edu/abs/2018MNRAS.476.4315N} {476, 4315}

\bibitem[\protect\citeauthoryear{{Nissanke}, {Vallisneri}, {Nelemans}  \&
  {Prince}}{{Nissanke} et~al.}{2012}]{2012ApJ...758..131N}
{Nissanke} S.,  {Vallisneri} M.,  {Nelemans} G.,   {Prince} T.~A.,  2012,
  \mn@doi [\apj] {10.1088/0004-637X/758/2/131}, \href
  {https://ui.adsabs.harvard.edu/abs/2012ApJ...758..131N} {758, 131}

\bibitem[\protect\citeauthoryear{{{\"O}zel}, {Psaltis}, {Narayan}  \& {Santos
  Villarreal}}{{{\"O}zel} et~al.}{2012}]{2012ApJ...757...55O}
{{\"O}zel} F.,  {Psaltis} D.,  {Narayan} R.,   {Santos Villarreal} A.,  2012,
  \mn@doi [\apj] {10.1088/0004-637X/757/1/55}, \href
  {https://ui.adsabs.harvard.edu/abs/2012ApJ...757...55O} {757, 55}

\bibitem[\protect\citeauthoryear{{Peters}}{{Peters}}{1964}]{1964PhRv..136.1224P}
{Peters} P.~C.,  1964, \mn@doi [Physical Review] {10.1103/PhysRev.136.B1224},
  \href {https://ui.adsabs.harvard.edu/abs/1964PhRv..136.1224P} {136, 1224}

\bibitem[\protect\citeauthoryear{{Peters} \& {Mathews}}{{Peters} \&
  {Mathews}}{1963}]{1963PhRv..131..435P}
{Peters} P.~C.,  {Mathews} J.,  1963, \mn@doi [Physical Review]
  {10.1103/PhysRev.131.435}, \href
  {https://ui.adsabs.harvard.edu/abs/1963PhRv..131..435P} {131, 435}

\bibitem[\protect\citeauthoryear{{Pratten}, {Klein}, {Moore}, {Middleton},
  {Steinle}, {Schmidt}  \& {Vecchio}}{{Pratten}
  et~al.}{2023}]{2023PhRvD.107l3026P}
{Pratten} G.,  {Klein} A.,  {Moore} C.~J.,  {Middleton} H.,  {Steinle} N.,
  {Schmidt} P.,   {Vecchio} A.,  2023, \mn@doi [\prd]
  {10.1103/PhysRevD.107.123026}, \href
  {https://ui.adsabs.harvard.edu/abs/2023PhRvD.107l3026P} {107, 123026}

\bibitem[\protect\citeauthoryear{{Roebber} et~al.,}{{Roebber}
  et~al.}{2020}]{2020ApJ...894L..15R}
{Roebber} E.,  et~al., 2020, \mn@doi [\apjl] {10.3847/2041-8213/ab8ac9}, \href
  {https://ui.adsabs.harvard.edu/abs/2020ApJ...894L..15R} {894, L15}

\bibitem[\protect\citeauthoryear{{Ruiter}, {Belczynski}, {Benacquista},
  {Larson}  \& {Williams}}{{Ruiter} et~al.}{2010}]{2010ApJ...717.1006R}
{Ruiter} A.~J.,  {Belczynski} K.,  {Benacquista} M.,  {Larson} S.~L.,
  {Williams} G.,  2010, \mn@doi [\apj] {10.1088/0004-637X/717/2/1006}, \href
  {https://ui.adsabs.harvard.edu/abs/2010ApJ...717.1006R} {717, 1006}

\bibitem[\protect\citeauthoryear{{Seto}}{{Seto}}{2001}]{2001PhRvL..87y1101S}
{Seto} N.,  2001, \mn@doi [\prl] {10.1103/PhysRevLett.87.251101}, \href
  {https://ui.adsabs.harvard.edu/abs/2001PhRvL..87y1101S} {87, 251101}

\bibitem[\protect\citeauthoryear{{Skilling}}{{Skilling}}{2004}]{2004AIPC..735..395S}
{Skilling} J.,  2004, in {Fischer} R.,  {Preuss} R.,   {Toussaint} U.~V.,  eds,
   AIP Conf. Series Vol. 735, AIP Conf. Series. pp 395--405,
  \mn@doi{10.1063/1.1835238}

\bibitem[\protect\citeauthoryear{{Skilling}}{{Skilling}}{2006}]{Skilling:2006gxv}
{Skilling} J.,  2006, \mn@doi [Bayesian Analysis] {10.1214/06-BA127}, 1, 833

\bibitem[\protect\citeauthoryear{{Tauris} \& {Sennels}}{{Tauris} \&
  {Sennels}}{2000}]{2000A&A...355..236T}
{Tauris} T.~M.,  {Sennels} T.,  2000, \mn@doi [\aap]
  {10.48550/arXiv.astro-ph/9909149}, \href
  {https://ui.adsabs.harvard.edu/abs/2000A&A...355..236T} {355, 236}

\bibitem[\protect\citeauthoryear{{Temmink}, {Toonen}, {Zapartas}, {Justham}  \&
  {G{\"a}nsicke}}{{Temmink} et~al.}{2020}]{2020A&A...636A..31T}
{Temmink} K.~D.,  {Toonen} S.,  {Zapartas} E.,  {Justham} S.,   {G{\"a}nsicke}
  B.~T.,  2020, \mn@doi [\aap] {10.1051/0004-6361/201936889}, \href
  {https://ui.adsabs.harvard.edu/abs/2020A&A...636A..31T} {636, A31}

\bibitem[\protect\citeauthoryear{{Tinto} \& {Dhurandhar}}{{Tinto} \&
  {Dhurandhar}}{2021}]{2021LRR....24....1T}
{Tinto} M.,  {Dhurandhar} S.~V.,  2021, \mn@doi [Living Reviews in Relativity]
  {10.1007/s41114-020-00029-6}, \href
  {https://ui.adsabs.harvard.edu/abs/2021LRR....24....1T} {24, 1}

\bibitem[\protect\citeauthoryear{{Toonen}, {Perets}, {Igoshev}, {Michaely}  \&
  {Zenati}}{{Toonen} et~al.}{2018}]{2018A&A...619A..53T}
{Toonen} S.,  {Perets} H.~B.,  {Igoshev} A.~P.,  {Michaely} E.,   {Zenati} Y.,
  2018, \mn@doi [\aap] {10.1051/0004-6361/201833164}, \href
  {https://ui.adsabs.harvard.edu/abs/2018A&A...619A..53T} {619, A53}

\bibitem[\protect\citeauthoryear{{Tutukov} \& {Yungelson}}{{Tutukov} \&
  {Yungelson}}{1993}]{1993MNRAS.260..675T}
{Tutukov} A.~V.,  {Yungelson} L.~R.,  1993, \mn@doi [\mnras]
  {10.1093/mnras/260.3.675}, \href
  {https://ui.adsabs.harvard.edu/abs/1993MNRAS.260..675T} {260, 675}

\bibitem[\protect\citeauthoryear{{Wagg}, {Broekgaarden}, {de Mink}, {Frankel},
  {van Son}  \& {Justham}}{{Wagg} et~al.}{2022}]{2022ApJ...937..118W}
{Wagg} T.,  {Broekgaarden} F.~S.,  {de Mink} S.~E.,  {Frankel} N.,  {van Son}
  L.~A.~C.,   {Justham} S.,  2022, \mn@doi [\apj] {10.3847/1538-4357/ac8675},
  \href {https://ui.adsabs.harvard.edu/abs/2022ApJ...937..118W} {937, 118}

\bibitem[\protect\citeauthoryear{Williams}{Williams}{2021}]{nessai}
Williams M.~J.,  2021, nessai: Nested Sampling with Artificial Intelligence,
  \mn@doi{10.5281/zenodo.4550693}, \url
  {https://doi.org/10.5281/zenodo.4550693}

\bibitem[\protect\citeauthoryear{{Williams}, {Veitch}  \&
  {Messenger}}{{Williams} et~al.}{2021}]{2021PhRvD.103j3006W}
{Williams} M.~J.,  {Veitch} J.,   {Messenger} C.,  2021, \mn@doi [\prd]
  {10.1103/PhysRevD.103.103006}, \href
  {https://ui.adsabs.harvard.edu/abs/2021PhRvD.103j3006W} {103, 103006}

\bibitem[\protect\citeauthoryear{{Zenati}, {Perets}  \& {Toonen}}{{Zenati}
  et~al.}{2019}]{2019MNRAS.486.1805Z}
{Zenati} Y.,  {Perets} H.~B.,   {Toonen} S.,  2019, \mn@doi [\mnras]
  {10.1093/mnras/stz316}, \href
  {https://ui.adsabs.harvard.edu/abs/2019MNRAS.486.1805Z} {486, 1805}

\bibitem[\protect\citeauthoryear{{van Kerkwijk} \& {Kulkarni}}{{van Kerkwijk}
  \& {Kulkarni}}{1999}]{1999ApJ...516L..25V}
{van Kerkwijk} M.~H.,  {Kulkarni} S.~R.,  1999, \mn@doi [\apjl]
  {10.1086/311991}, \href
  {https://ui.adsabs.harvard.edu/abs/1999ApJ...516L..25V} {516, L25}

\makeatother
\end{thebibliography}



\appendix

\section{The dependence of eccentricity measurability on other source properties}\label{app:vary_Tobs}

The ability of \LISA to measure (or place upper limits on) the eccentricity of quasi-monochromatic binaries naturally depends on the source parameters.
The minimum detectable eccentricity depends sensitively on the GW frequency of the source and the SNR; this was quantified in Sec.~\ref{sec:results}.
However, it is not expected to depend sensitively on any of the other source or mission parameters provided they are varied in such a way as the SNR is kept constant.

Firstly, the dependence of the minimum detectable eccentricity on the source chirp rate, $\dot{f}_{\rm GW}$, is investigated.
The fiducial source indicated by the circular marker in Fig.~\ref{fig:min_ecc} was reanalysed varying the injected $\dot{f}_{\rm GW}$ while keeping all other parameters fixed. 
The GR value of $\dot{f}_{\rm GW}$ (Eq.~\ref{eq:chirp}) was modified by a factor $\alpha$ in the range $\alpha\in[0.8,1.2]$.
Nine values of $\alpha$ regularly spaced in this range were investigated and all the 1-dimensional marginalised posteriors on the eccentricity parameter appear indistinguishable to the eye. 
Quantitatively, the Kullback–Leibler (KL) divergence \citep{MacKay2003} between the posterior obtained with the GR value of $\dot{f}_{\rm GW}$ (i.e.\ $\alpha=1$) and the posterior with a general $\alpha$ was less than $2\times 10^{-3}$ in all cases.

Secondly, the dependence of the minimum detectable eccentricity on the sky position was investigated.
The fiducial source was reanalysed with 10 randomised sky locations, while keeping all other parameters fixed. 
For each sky new position the strain amplitude was adjusted slightly in order to keep the SNR constant.
Again, all the 1-dimensional marginalised posteriors on the eccentricity parameter appear indistinguishable to the eye and the KL divergence between the posterior of the fiducial source and that for the run with the randomised sky position was less than $2\times 10^{-3}$ in all cases.

Thirdly, the dependence of the minimum detectable eccentricity on the source inclination was investigated.
The fiducial source was reanalysed with 7 different inclinations uniformly spaced in cosine inclination in the range [-1,+1], while keeping all other parameters fixed and adjusting the amplitude to keep the SNR constant.
Again, all the 1-dimensional marginalised posteriors on the eccentricity parameter appear indistinguishable to the eye and the KL divergence between the posterior of the fiducial source $\lesssim 2\times 10^{-3}$ in all cases.

\begin{figure}
    \centering
    \includegraphics[width=0.99\columnwidth]{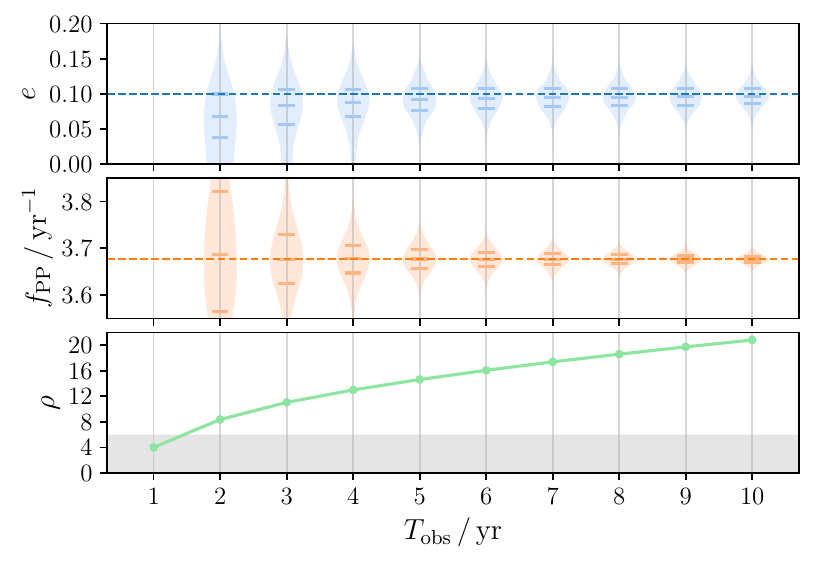}
    \caption{
        The improvement in eccentricity measurement with increasing mission duration, $T_{\rm obs}$.
        \emph{Top panel:} posteriors on the eccentricity.
        \emph{Middle panel:} the posteriors on the periastron precession frequency.
        Blue (orange) horizontal lines indicate the injected parameter values and the median and 50\% credible intervals are indicated in each violin plot.
        The eccentricity can be confidently measured to be greater than zero for all $T_{\rm obs}>4$ years and the posterior widths on both $e$ and $f_{\rm PP}$ scale as $\rho^{-1}$ for longer mission durations.
        \emph{Bottom panel:} the total signal SNR. 
        For a 1 year mission the SNR is below 6 (gray shaded region) and the source cannot be confidently detected (this is why no posteriors on $e$ or $f_{\rm PP}$ are plotted for $T_{\rm obs}=1$ year).
    }
    \label{fig:ecc_post_Tobs}
\end{figure}

Finally, the dependence of the minimum detectable eccentricity on the mission duration was investigated. 
The fiducial source was reanalysed with simulated mission durations of 1, 2, \ldots, 10 years, see Fig.~\ref{fig:ecc_post_Tobs}. 
The SNR increases with time, leading to improved eccentricity measurements. 
For mission durations longer than a couple of years, the improvement in the eccentricity measurement has the expected scaling of $\rho^{-1}$, indicating that there is no other effect on the eccentricity measurement besides that of the increased SNR. 

Taken together, these results show that \LISA's ability to measure eccentricity in quasi-monochromatic binaries depends only very weakly on all the source and mission parameters except the source frequency, $f_{\rm GW}$, and the SNR, $\rho$. 
This justifies the use of the fitting formula in Eq.~\ref{eq:analytic_emin} that is a function of just these two parameters.

\section{Using the Bayes' factor to quantify the measurability of eccentricity}\label{app:BF}

\begin{figure*}
    \centering
    \begin{subfigure}[b]{0.98\textwidth}
        \includegraphics[width=0.98\textwidth]{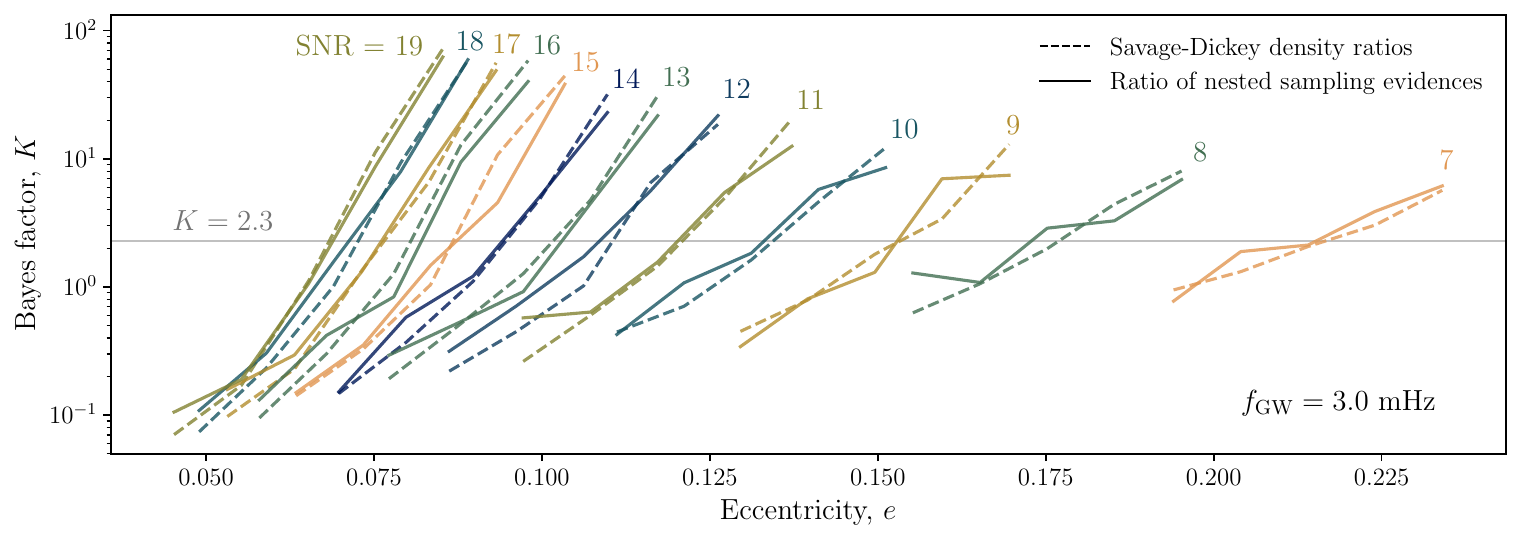}
    \end{subfigure}
    \begin{subfigure}[b]{0.98\textwidth}
        \includegraphics[width=0.98\textwidth]{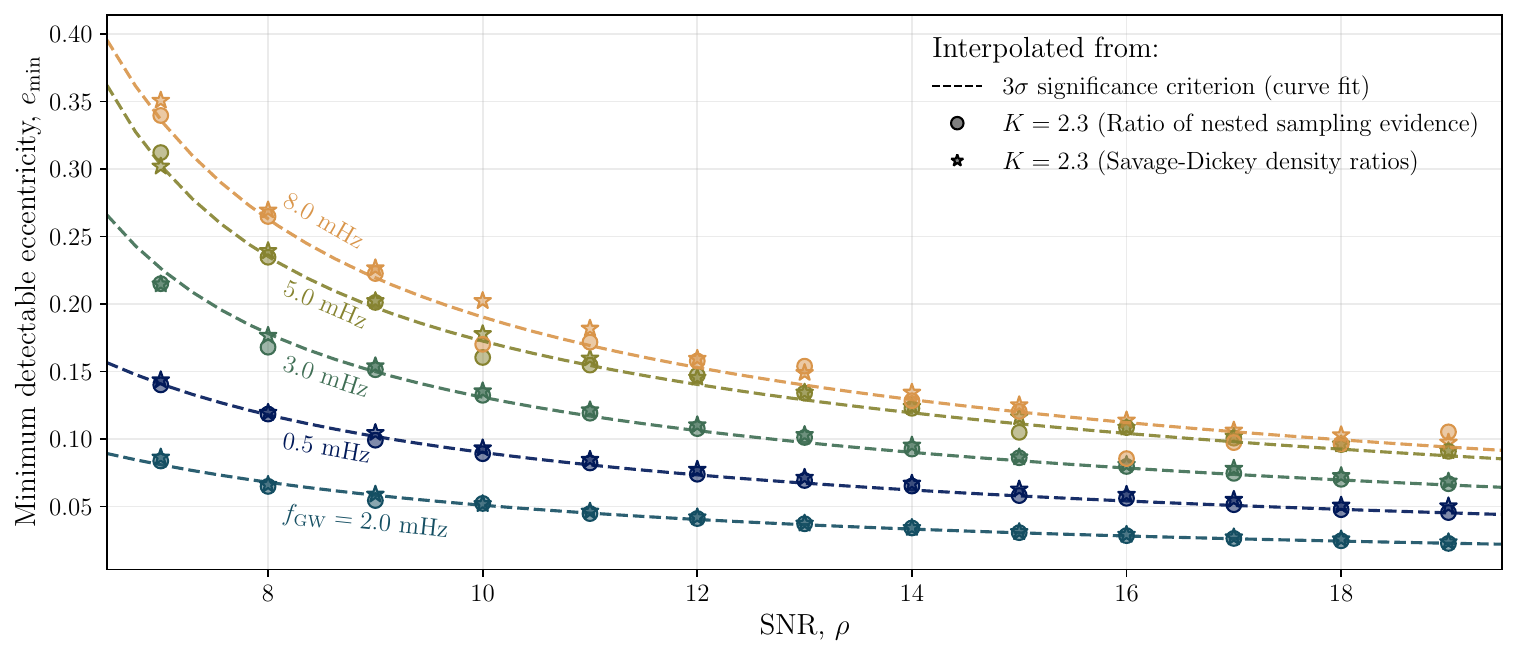}
    \end{subfigure}
    \caption{ \label{fig:BayesFactorCheck}
        The bottom panel shows the minimum detectable eccentricity as a function of SNR, for a subset of the GW frequencies tested.
        Dashed lines show the analytic fit in Eq.~\ref{eq:analytic_emin} to the data in Fig.~\ref{fig:min_ecc} obtained using a threshold on the median over standard deviation of the posterior on $e$.
        Circular (star-shaped) markers show the results obtained using a threshold on the Bayes' factor calculated using the nested sampling (Savage-Dickey) method.
        These were obtained from a sequence of 5 runs with increasing eccentricity (at fixed values of all other parameters), computing the Bayes' factor for each run, linearly interpolating the Bayes' factor as a function of eccentricity, and determining the value of $e_{\rm min}$ where the Bayes' factor exceeds a threshold $K$. 
        The top panel shows the Bayes' factors as a function of eccentricity for an example sequence of runs at $f_{\rm GW}=3\,\mathrm{mHz}$.
        The threshold on the Bayes' factor of $K=2.3$ was determined by minimising the root-mean-square (r.m.s.) difference between the circular points and the dashed lines, the resulting r.m.s.\ difference is $\Delta e_{\rm min}=0.006$. 
        If instead the r.m.s. difference between the Savage-Dickey results (star-shaped points) and the dashed lines is minimised, a threshold of $K=1.9$ was obtained, with an r.m.s.\ difference of $\Delta e_{\rm min}=0.004$.
        Overall, the Bayes' factor results agree well with the results in Fig.~\ref{fig:min_ecc}, justifying the use of the detection threshold used.
    }
\end{figure*}

The results for the minimum detectable eccentricity, $e_{\rm min}(f_{\rm GW}, \rho)$, in Fig.~\ref{fig:min_ecc} were obtained by requiring that the 1-dimensional marginalised posterior on the eccentricity parameter was peaked sufficiently away from zero. 
This was quantified using the ratio of the median to the standard deviation of the posterior distribution on $e$ and this ratio was required to be greater than three in order to claim the eccentricity as measurable.

While this threshold for eccentricity detection has the advantage of being simple to implement and simple to interpret visually (see inset in Fig.~\ref{fig:min_ecc}) it is also somewhat arbitrary and difficult to relate to other statistical measures of detection confidence.
Here we reproduce similar results for $e_{\rm min}$ using a threshold on Bayes' factor.
The Bayes' factor is defined as the ratio of the Bayesian evidence between an analysis using an eccentric waveform model and an analysis using a circular model ($e=0$).
An equal prior odds ratio was used between these two models.
The Bayes' factor is calculated in two ways: using the ratio of the two evidences calculated via nested sampling and (exploiting the fact that the circular model is nested in the eccentric model) using the Savage-Dickey density ratio on the parameter $e$.

Using the Bayes' factor as a statistic for detecting the eccentricity, we now claim a measurement of the eccentricity if the Bayes' factor exceeds a certain predetermined threshold, $K$.
The results are shown in Fig.~\ref{fig:BayesFactorCheck}.
This analysis gives an independent check on the results in Fig.~\ref{fig:min_ecc} and allows us to establish that our previous threshold of three on the ratio of median to the standard deviation corresponds to a threshold on the Bayes' factor of $K = 2.3$.

\section{The effect of including subdominant harmonics} \label{app:varyN}

Most of the parameter estimation calculations performed so far in this paper have either been circular analyses targeting a single harmonic, or eccentric analyses targeting all $N=12$ harmonics. 
This appendix investigates eccentric analyses that analyse only a smaller number of harmonics with $N$ in the range 2 to 12.

The benefit of analysing a small number of harmonics is increased computation efficiency. 
The disadvantage is the loss of SNR when only analysing a limited number of harmonics. 
Although this is mitigated by the fact that the SNR is concentrated in a small number of harmonics for the small to moderated eccentricities expected for Galactic WD+NS binaries (see Table \ref{tab:D}). 

\begin{figure*}
    \centering
    \includegraphics[width=0.98\textwidth]{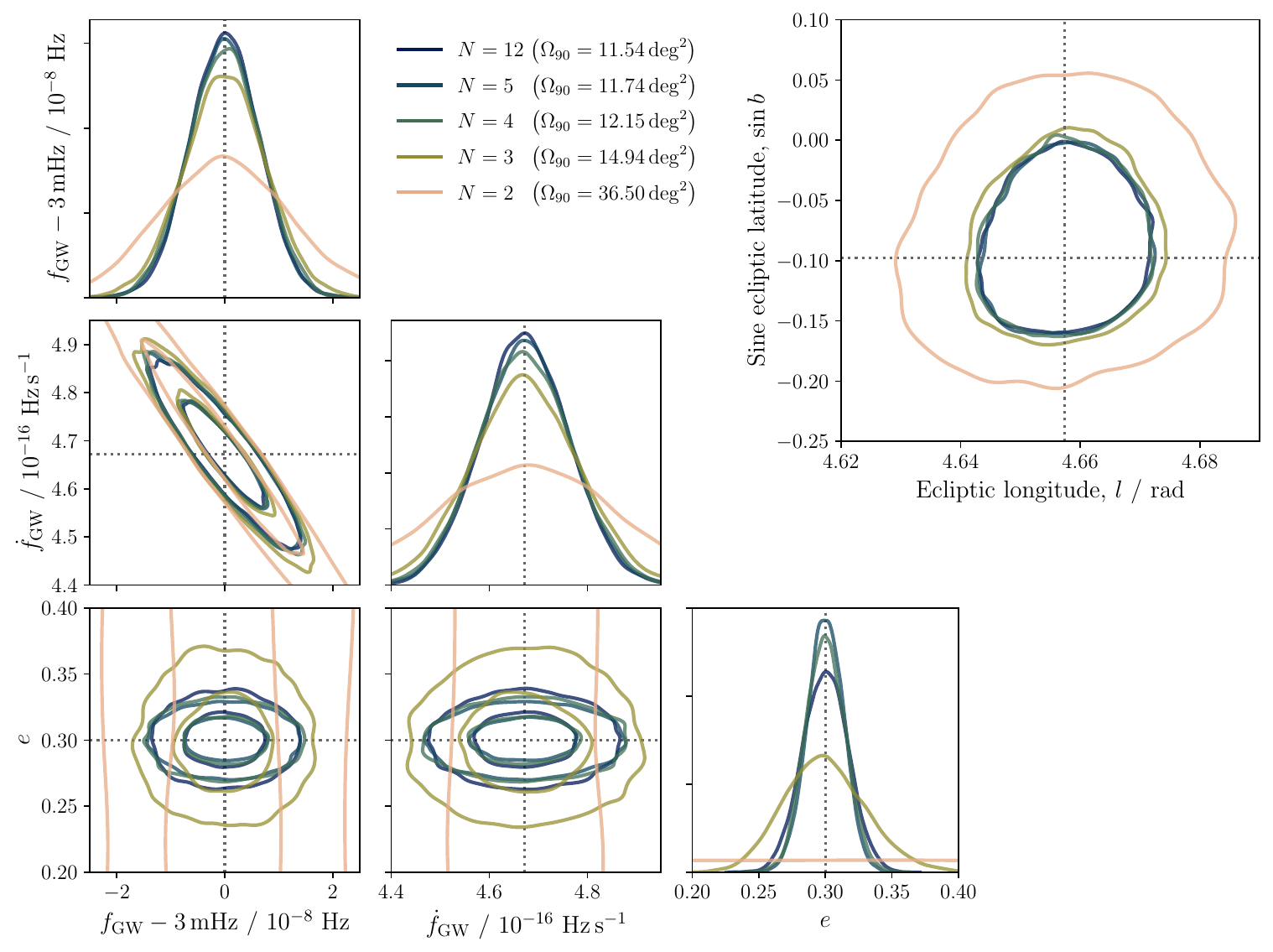}
    \caption{ \label{fig:varyN}
        Posterior distributions on selected parameters for an eccentric source analysed including different numbers of harmonics, $N$.
        Including more harmonics leads to improved parameter constraints, if it has any noticeable effect at all.
        The biggest improvement comes in the eccentricity parameter when going from $N=2$ to 3; at least 2 usable (loud) harmonics are needed to constrain $e$ (and the $n=1$ contains very little SNR).
        In all other cases a significant improvement is found when going from $N=2$ to 3 and a very small but measurable improvement is found when going from $N=3$ to 4 (particularly in the sky localisation). 
        Beyond $N=4$ no improvements are found. 
        These results can be understood in terms of the harmonic SNRs in Table \ref{tab:D}.
        Dotted vertical and horizontal lines indicate the injected source parameters which are the same as those used in Fig.~\ref{fig:circ_ecc_PE}.
        In the 2-dimensional posteriors in the corner plot on the left-hand side 50\% and 90\% contours are shown, whereas in the sky map on the right-hand side only the 90\% contours are shown for clarity.
        Areas of the 90\% sky regions, $\Omega_{90}$, are also given in the legend.
    }
\end{figure*}

The eccentric source used in Fig.~\ref{fig:circ_ecc_PE} was reanalysed using a different number of total harmonics. 
The results are shown in Fig.~\ref{fig:varyN}.
As expected, the effect of including a particular harmonic scale with the SNR in that harmonic.
However, even harmonics that contain small SNR (significantly below the threshold of $\sim 6$ that would be required for them to be identified as an individual source in a circular source search) still have a noticeable effect on the posterior.

\bsp	
\label{lastpage}
\end{document}